\documentclass[]{imaiai}
\usepackage{amssymb}
\usepackage{placeins}
\usepackage{multirow}
\usepackage{graphicx}
\usepackage[caption=false]{subfig} 
\usepackage{pifont}
\usepackage{anyfontsize} 
\usepackage{tabularx} 
\usepackage{url}
\usepackage{navigator} 
\usepackage{amsmath} 
\usepackage{amsthm} 
\usepackage{epstopdf} 		
\DeclareMathOperator{\Tr}{Tr} 

\usepackage{xcolor}
\usepackage{soul}

\begin{document}
	
	\title{Balance and Frustration in Signed Networks}

	\shorttitle{Balance and Frustration in Signed Networks} 
	\shortauthorlist{S. Aref and M.C. Wilson} 
	
	\author{{
			\sc Samin Aref}$^{1,2}$
		{\sc and}
		{\sc Mark C. Wilson}$^{1}$
		\\[2pt]
		$^{1}$
		Department of Computer Science, University of Auckland\\
		Auckland, Private Bag 92019, New Zealand\\
		$^{2}$Laboratory of Digital and Computational Demography, Max Planck Institute for Demographic Research, Konrad-Zuse-Str. 1, 18057 Rostock, Germany\\
		{\email{sare618@aucklanduni.ac.nz}}
	}

\maketitle

\begin{abstract}
{The frustration index is a key measure for analysing signed networks, which has been underused due to its computational complexity. We use an exact optimisation-based method to analyse frustration as a global structural property of signed networks coming from diverse application areas. In the classic friend-enemy interpretation of balance theory, a by-product of computing the frustration index is the partitioning of nodes into two internally solidary but mutually hostile groups.
	
The main purpose of this paper is to present general methodology for answering questions related to partial balance in signed networks, and apply it to a range of representative examples that are now analysable because of advances in computational methods.

We provide exact numerical results on social and biological signed networks, networks of formal alliances and antagonisms between countries, and financial portfolio networks. Molecular graphs of carbon and Ising models are also considered. The purpose served by exploring several problems in this paper is to propose a single general methodology for studying signed networks and to demonstrate its relevance to applications. We point out several mistakes in the signed networks literature caused by inaccurate computation, implementation errors or inappropriate measures.}
{Frustration index,
	Line index of balance,
	Signed graph,
	Integer programming,
	Optimisation,
	Balance theory}

Mathematics Subject Classification (MSC 2010): 05C22, 90C35, 90C90, 90C09, 90C10

\footnotetext{The reference to this article should be made as follows: {\scshape Aref, S., and Wilson, M.~C.}
	\newblock Balance and frustration in signed networks.
	\newblock {\em Journal of Complex Networks 7}, 2 (2019), 163--189.
	\newblock doi: 10.1093/comnet/cny015.}
\end{abstract}

\section{Introduction} \label{s:intro}

The theory of structural balance introduced by Heider \cite{heider_social_1944} is an essential tool in the context of social relations for understanding the impact of local interactions on the global structure of signed networks. Following Heider, Cartwright and Harary identified cycles containing an odd number of negative edges \cite{cartwright_structural_1956} as a source of tension that may influence the structure of signed networks in particular ways. Signed networks in which no such cycles are present satisfy the property of structural balance, which is considered as a state with minimum tension \cite{cartwright_structural_1956}.

For graphs that are not balanced, a distance from balance (a measure of partial balance) can be computed \cite{aref2015measuring}. In the past few decades, different measures of balance have been suggested. Measures based on cycles \cite{cartwright_structural_1956, harary_structural_1957, harary_measurement_1959, norman_derivation_1972, harary1977graphing}, triangles \cite{terzi_spectral_2011,kunegis_applications_2014}, and eigenvalues of the adjacency matrix \cite{kunegis_spectral_2010, pelino2012towards, estrada_walk-based_2014, kunegis_applications_2014} are not generally consistent \cite{aref2015measuring} and result in conflicting observations \cite{leskovec_signed_2010, facchetti_computing_2011, estrada_walk-based_2014}. A recent study \cite{aref2015measuring} compares all these measures and concludes that substantially different levels of partial balance are observed under cycle-based \cite{cartwright_structural_1956, harary_structural_1957, harary_measurement_1959, norman_derivation_1972, harary1977graphing, terzi_spectral_2011, kunegis_applications_2014}, eigenvalue-based \cite{kunegis_spectral_2010, pelino2012towards, estrada_walk-based_2014, kunegis_applications_2014}, and frustration-based \cite{abelson_symbolic_1958, harary_measurement_1959, zaslavsky_balanced_1987, facchetti_computing_2011} measures of balance.

Among various measures is the \textit{frustration index} that indicates the minimum number of edges whose removal (or equivalently, negation) results in balance \cite{abelson_symbolic_1958,harary_measurement_1959,zaslavsky_balanced_1987}. The clear definition of the frustration index allows for an intuitive interpretation of its values as the minimum number of edges that keep the network away from a state of total balance (an edge-based distance from balance). In this study, we focus on applications of the frustration index, also known as \textit{the line index of balance} \cite{harary_measurement_1959}, in different contexts beyond the structural balance of signed social networks. 

Satisfying essential axiomatic properties as a measure of partial balance \cite{aref2015measuring}, the frustration index is a key to frequently stated problems in many different fields of research \cite{iacono_determining_2010,kasteleyn_dimer_1963,patrick_doreian_structural_2015,doslic_computing_2007}. In biological networks, optimal decomposition of a network into monotone subsystems 
is made possible by calculating the frustration index of the underlying signed graph \cite{iacono_determining_2010}. In physics, the frustration index provides the ground state of atomic magnet models \cite{kasteleyn_dimer_1963,hartmann2011ground}. In international relations, the dynamics of alliances and enmities between countries can be investigated using the frustration index \cite{patrick_doreian_structural_2015}. Frustration index can also be used as an indicator of network bi-polarisation in practical examples involving financial portfolios. For instance, some low-risk portfolios are shown to have an underlying balanced signed graph containing negative edges \cite{harary_signed_2002}. In chemistry, bipartite edge frustration can be used as a stability indicator of carbon allotropes known as fullerenes \cite{doslic2005bipartivity, doslic_computing_2007}. 

\section{Computing the frustration index} \label{s:computing}
From a computational viewpoint, computing the frustration index of a signed graph is an NP-hard problem equivalent to the ground state calculation of an Ising model without special structure \cite{Sherrington, Barahona1982,mezard2001bethe}. Computation of the frustration index also reduces from classic unsigned graph optimisation problems (EDGE-BIPARTIZATION and MAXCUT) which are known to be NP-hard \cite{huffner_separator-based_2010}.

The frustration index can be computed in polynomial time for planar graphs \cite{katai1978studies,huffner_separator-based_2010}. In general graphs; however, the frustration index is even NP-hard to approximate within any constant factor \cite{huffner_separator-based_2010}. 
There has been a lack of systematic investigations for computing the exact frustration index of large-scale networks \cite{aref2016exact,aref2017computing}. In small graphs with fewer than 40 nodes, exact computational methods \cite{flament1963applications,hammer1977pseudo,bramsen2002further,brusco_k-balance_2010} are used to obtain the frustration index. More recent studies focus on approximating \cite{dasgupta_algorithmic_2007,agarwal2005log,avidor2007multi,coleman2008local} and computing \cite{aref2016exact, aref2017computing} the frustration index in large signed graphs with at least thousands of edges.


A closely related and more general problem (that is beyond our discussions in this paper) is finding the minimum number of edges whose removal results in a weakly balanced signed graph (as in Davis's definition of \textit{weak balance} \cite{Davis}). This problem is referred to as the \textit{Correlation Clustering} problem \cite{figueiredo2013mixed, ma_memetic_2015} which is investigated more comprehensively in the literature \cite{Giotis,brusco_k-balance_2010, figueiredo2013mixed,drummond2013efficient,levorato2015ils,levorato2017evaluating}. A comparison of mathematical programming models for computing the frustration index and correlation clustering can be found in \cite[Subsection 8.2]{aref2016exact}.


Facchetti, Iacono, and Altafini suggested a non-linear energy function minimisation model for finding the frustration index \cite{facchetti_computing_2011}. Their model was used as the basis of various non-exact optimisation techniques \cite{iacono_determining_2010, esmailian_mesoscopic_2014, ma_memetic_2015, ma_decomposition-based_2017, Wang2016}. Using heuristic algorithms \cite{iacono_determining_2010}, estimations of the frustration index have been provided for biological networks up to $1.5\times10^3$ nodes \cite{iacono_determining_2010} and social networks with up to $10^5$ nodes \cite{facchetti_computing_2011, facchetti2012exploring}. 
Doreian and Mrvar \cite{patrick_doreian_structural_2015} have provided some upper bounds on the frustration index of signed international relation networks \cite{correlatesofwar2004}. We use their dataset in Section~\ref{s:temporal} and analyse it using the exact values of the frustration index.




In this study, we use a recently developed optimisation model \cite{aref2016exact} that computes the frustration index of large-scale signed networks exactly and efficiently.

\subsection*{\textbf{Our contribution}} \label{ss:contrib}

We focus on the frustration index of signed networks, a standard measure of balance mostly estimated or approximated for decades due to the inherent combinatorial complexity. We continue a line of research begun in \cite{aref2015measuring} (which compared various measures of partial balance and suggested that the frustration index should be more widely used), \cite{aref2017computing} (which explained how integer linear optimisation models can be used to compute the frustration index) and \cite{aref2016exact} (which substantially improves the efficiency of such computations using algorithmic refinements and powerful mathematical programming solvers).

The purpose of the present article is to present a single general methodology for studying signed networks and to demonstrate its relevance to applications. We consider a variety of signed networks arising from several disciplines. These networks differ substantially in size and the computational results require different interpretations. The current implementation of our algorithms can efficiently provide exact results on networks with up to 100000 edges. A by-product of exactly computing the frustration index is an optimal partitioning of nodes into two groups where the number of intra-group negative edges and inter-group positive edges is minimised.


This paper begins by laying out the theoretical dimensions of the research in Section~\ref{s:prelim}. The computational method is briefly discussed in Section~\ref{s:model} followed by a discussion on its efficiency. 
Numerical results on signed networks of six disciplines are provided in Sections \ref{s:d1} -- \ref{s:d56}. Section~\ref{s:conclu} provides a short conclusion and suggests future directions.
Along the way we point out several mistakes in the signed networks literature caused by inappropriate measures and inaccurate computation.

\section{Preliminaries} \label{s:prelim}

We recall some standard definitions.

\subsection{Notation} 
We consider undirected signed networks $G = (V,E,\sigma)$. The ordered set of nodes is denoted by $V=\{1, 2, \dots, n\}$, with $|V| = n$. The set $E$ of edges can be partitioned into the set of positive edges $E^+$ and the set of negative edges $E^-$ with $|E|=m$, $|E^-|=m^-$, and $|E^+|=m^+$ where $m=m^- + m^+$. The sign function is denoted by $\sigma: E\rightarrow\{-1,+1\}$.

We represent the $m$ undirected edges in $G$ as ordered pairs of vertices $E = \{e_1, e_2, \dots , e_m\} \subseteq \{ (i,j) \mid i,j \in V , i<j \}$, where a single edge between nodes $i$ and $j$, $i<j$, is denoted by $(i,j) , i<j$. We denote the graph density by $\rho= 2m/(n(n-1))$.

The entries of the symmetric adjacency matrix $\textbf{A}$ are defined in Eq.\ \eqref{eq1}. 
\begin{equation}\label{eq1}
a{_i}{_j} =
\left\{
\begin{array}{ll}
\sigma_{(i,j)} & \mbox{if } (i,j) \in E \\
\sigma_{(j,i)} & \mbox{if } (j,i) \in E \\
0 &  \text{otherwise}
\end{array}
\right.
\end{equation}



We use $G_r=(V,E,\sigma_r)$ to denote a reshuffled graph in which the sign function $\sigma_r$ is a random mapping of $E$ to $\{-1,+1\}$ that preserves the number of negative edges. 

A \emph{walk} of length $k$ in $G$ is a sequence of nodes $v_0,v_1,...,v_{k-1},v_k$ such that for each $i=1,2,...,k$ there is an edge from $v_{i-1}$ to $v_i$. If $v_0=v_k$, the sequence is a \emph{closed walk} of length $k$. If the nodes in a closed walk are distinct except for the endpoints, the walk is a cycle of length $k$. The \emph{sign} of a walk or cycle is the product of the signs of its edges. Cycles with positive (negative) signs are balanced (unbalanced). A balanced graph is one with no unbalanced cycles.

\subsection{Frustration count}


For any signed graph $G=(V, E, \sigma)$, we can partition $V$ into two sets, denoted $X \subseteq V$ and $\bar X=V \backslash X$. We call $X$ a colouring set and we think of this partitioning as specifying a colouring of the nodes, where each node $i \in X$ is coloured black, and each node $i \in \bar X$ is coloured white. We let $x_i$ denote the colour of node $i \in V$ under $X$, where $x_i=1$ if $i \in X$ and $x_i=0$ otherwise. Considering node colours is a standard way of defining discrete decision variables for nodes in modelling graph optimisation problems.

\begin{figure}[ht]
	\subfloat[An example signed graph]{\includegraphics[height=2in]{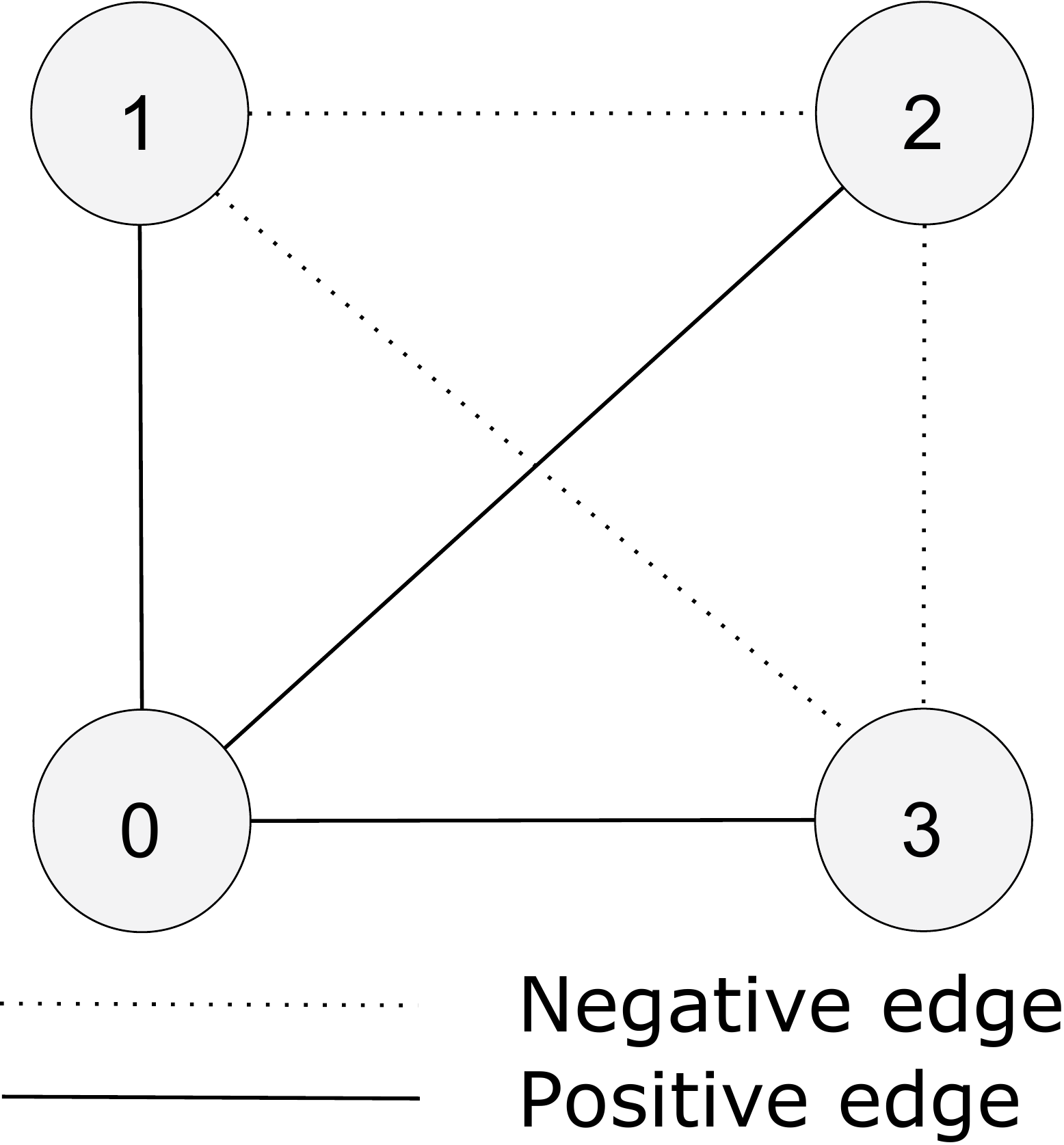}%
		\label{fig1a}}
	\hfil
	\subfloat[Two node colourings (both optimal) and their resulting frustrated edges]{\includegraphics[height=2in]{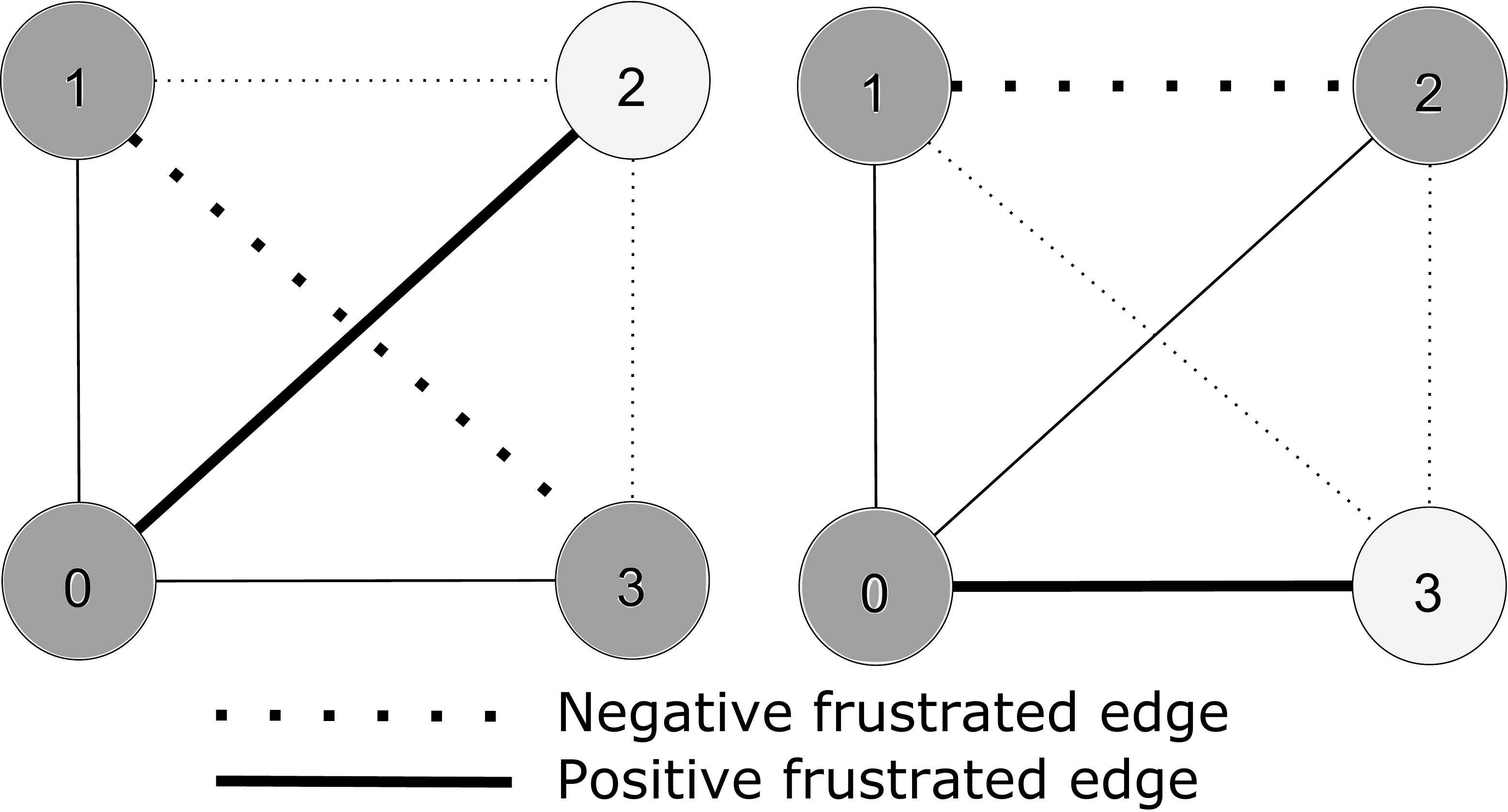}%
		\label{fig1b}}
	\hfil
	\caption{The impact of node colouring on the frustration of edges}
	\label{fig1}
\end{figure}

Figure \ref{fig1} (a) demonstrates an example signed graph in which positive and negative edges are represented by solid lines and dotted lines respectively. Figure~\ref{fig1} (b) illustrates two node colourings and the resulting frustrated edges represented by thick lines. 

We define the {\em frustration count} $f_G(X)$ as the number of frustrated edges of $G$ under $X$: $$f_G(X) = \sum_{(i,j) \in E} f_{ij}(X)$$
where $f_{ij}(X)$ is the frustration state of edge $(i,j)$, given by
\begin{equation} \label{eq2}
f_{ij}(X)=
\begin{cases}
0, & \text{if}\ x_i = x_j \text{ and } (i,j) \in E^+ \\
1, & \text{if}\ x_i = x_j \text{ and } (i,j) \in E^- \\
0, & \text{if}\ x_i \ne x_j \text{ and } (i,j) \in E^- \\
1, & \text{if}\ x_i \ne x_j \text{ and } (i,j) \in E^+ \\
\end{cases}
\end{equation}

The frustration index $L(G)$ of a graph $G$ can be obtained by finding a subset $X^* \subseteq V$ of $G$ that minimises the frustration count $f_G(X)$, i.e., solving Eq.\ \eqref{eq3}. 

\begin{equation} \label{eq3}
L(G) = \min_{X \subseteq V}f_G(X)\
\end{equation}


\section{Methods and Materials} \label{s:model}

In this section, we discuss a mathematical programming model in Eq.\ \eqref{eq4} that minimises the frustration count as the objective function over binary decision variables defined for nodes and edges of the graph. Note that, there are various ways to formulate the frustration count using variables defined over graph nodes and edges leading to various mathematical programming models that are investigated in detail in \cite{aref2016exact}.

\subsection{Methodology}

Computing the frustration index can be formulated as a binary linear model that counts the frustrated edges using binary variable $f_{ij} \: \forall (i,j) \in E$ to denote frustration of edge $(i,j)$ and $x_i \: \forall i \in V$ to denote the colour of node $i$. Constraints of the optimisation model are formulated according to the values of $f_{ij}$ in Eq.\ \eqref{eq2}. Therefore, the minimum frustration count under all node colourings is obtained by solving \eqref{eq4}:
\begin{equation}\label{eq4}
\begin{split}
\min_{x_i, f_{ij}} Z &= \sum\limits_{(i,j) \in E}  f_{ij}  \\
\text{s.t.} \quad
f_{ij} &\geq x_{i}-x_{j} \quad \forall (i,j) \in E^+ \\
f_{ij} &\geq x_{j}-x_{i} \quad \forall (i,j) \in E^+ \\
f_{ij}  &\ge  x_{i} + x_{j} -1 \quad \forall (i,j) \in E^- \\
f_{ij}  &\ge  1-x_{i} - x_{j}  \quad \forall (i,j) \in E^- \\
x_{i} &\in \{0,1\} \quad  \forall i \in V \\
f_{ij} &\in \{0,1\} \quad \forall (i,j) \in E 
\end{split}
\end{equation}

Several techniques are used to speed up the branch and bound algorithm for solving the binary programming model in Eq.\ \eqref{eq4} and the like as discussed in \cite{aref2016exact,aref2017computing}. We implement the three speed-up techniques known as pre-processing data reduction, prioritised branching and fixing a colour, and unbalanced triangle valid inequalities
 \cite{aref2016exact,aref2017computing} and solve the optimisation problem using \textit{Gurobi} \cite{gurobi}. Details of the optimisation model and speed-up techniques can be found in \cite{aref2016exact,aref2017computing}.


Aref et al.\ have tested their optimisation models \cite{aref2016exact,aref2017computing} on synthetic and real-world datasets using ordinary desktop computers showing the efficiency of their models in computing the frustration index in comparison to other models in the literature \cite{flament1963applications,hammer1977pseudo,hansen_labelling_1978,harary_simple_1980,bramsen2002further,dasgupta_algorithmic_2007, brusco_k-balance_2010, huffner_separator-based_2010, iacono_determining_2010,patrick_doreian_structural_2015}. For detailed discussions on the efficiency of the binary linear programming model in Eq.\ \eqref{eq4}, one may refer to \cite{aref2016exact}.

For comparing the level of frustration among networks of different size and order, we use the \textit{normalised frustration index}, $F(G)=1-2L(G)/m$. This standard measure of partial balance is suggested in a comparative study of structural balance measures because it satisfies key axiomatic properties \cite{aref2015measuring}. Values of $F(G)$ are within the range of $[0,1]$ and greater values of $F(G)$ represent closeness to a state of structural balance.

Our baseline for evaluating balance comprises the average and standard deviation of the frustration index in reshuffled graphs (that have signs allocated randomly to the same underlying structure). Accordingly, we use Z score values, $Z={(L(G)-L(G_r))}/{\text{SD}}$, in order to evaluate the level of partial balance precisely.

\subsection{Materials}

We use a wide range of examples from different disciplines all being undirected signed networks. This includes four social signed networks ranging in size from 49 to 99917 edges in Section~\ref{s:d1}, four biological signed networks with 779-3215 edges in Section~\ref{s:d2}, one dynamic network of international relations with 51 time windows ranging in size from 362 to 1247 edges in Section~\ref{s:d3}, six financial portfolios with 10-55 edges over 9 years in Section~\ref{s:d4}, and molecular fullerene graphs with 270-9000 edges and Ising models with 32-79600 edges in Section~\ref{s:d56}. The datasets used in this study are made publicly available on the \urllink{https://figshare.com/articles/Signed_networks_from_sociology_and_political_science_biology_international_relations_finance_and_computational_chemistry/5700832}{Figshare} research data repository \cite{Aref2017data}. We use a wide variety of datasets, rather than focusing on a specific application, in order to underline the generality of our approach.

The fundamental reason why we only use undirected signed networks is that the reliability test for predictions on directed signed networks made by balance theory shows very negative results \cite{leskovec_signed_2010}. Based on large directed signed networks such as Epinions, Slashdot, and Wikipedia, the binary predictions made by balance theory are incorrect almost half of the time \cite{leskovec_signed_2010}. This observation supports the inefficacy of balance theory for structural analysis of \emph{directed} signed graphs \cite{aref2015measuring}. 

The numerical results in this paper are obtained by solving the binary linear programming model \eqref{eq4} coupled with three speed-up techniques \cite{aref2016exact} using Gurobi's Python interface \cite{gurobi}. Unless stated otherwise, the hardware used for the computational analysis is a virtual machine with 32 Intel Xeon CPU E5-2698 v3 @ 2.30 GHz processors and 32 GB of RAM running 64-bit Microsoft Windows Server 2012 R2 Standard.

\section{Social networks}\label{s:d1}

In this section, we discuss using the frustration index to analyse social signed networks inferred from the sociology and political science datasets.

\subsection{Datasets}

We use well-studied datasets of communities with positive and negative interactions and preferences. This includes Read's dataset for New Guinean highland tribes \cite{read_cultures_1954} and the last time frame of Sampson's data on monastery interactions \cite{sampson_novitiate_1968}. 
Our analysis also includes a signed network of US senators that is inferred in \cite{neal_backbone_2014} through implementing a stochastic degree sequence model on Fowler's Senate bill co-sponsorship data \cite{fowler_legislative_2006} for the 108th US senate.

\begin{figure}[ht] 
	\centering
	\subfloat[Optimal colouring of New Guinean tribes network with frustrated edges shown by thick lines \cite{read_cultures_1954}]{\includegraphics[height=1.7in]{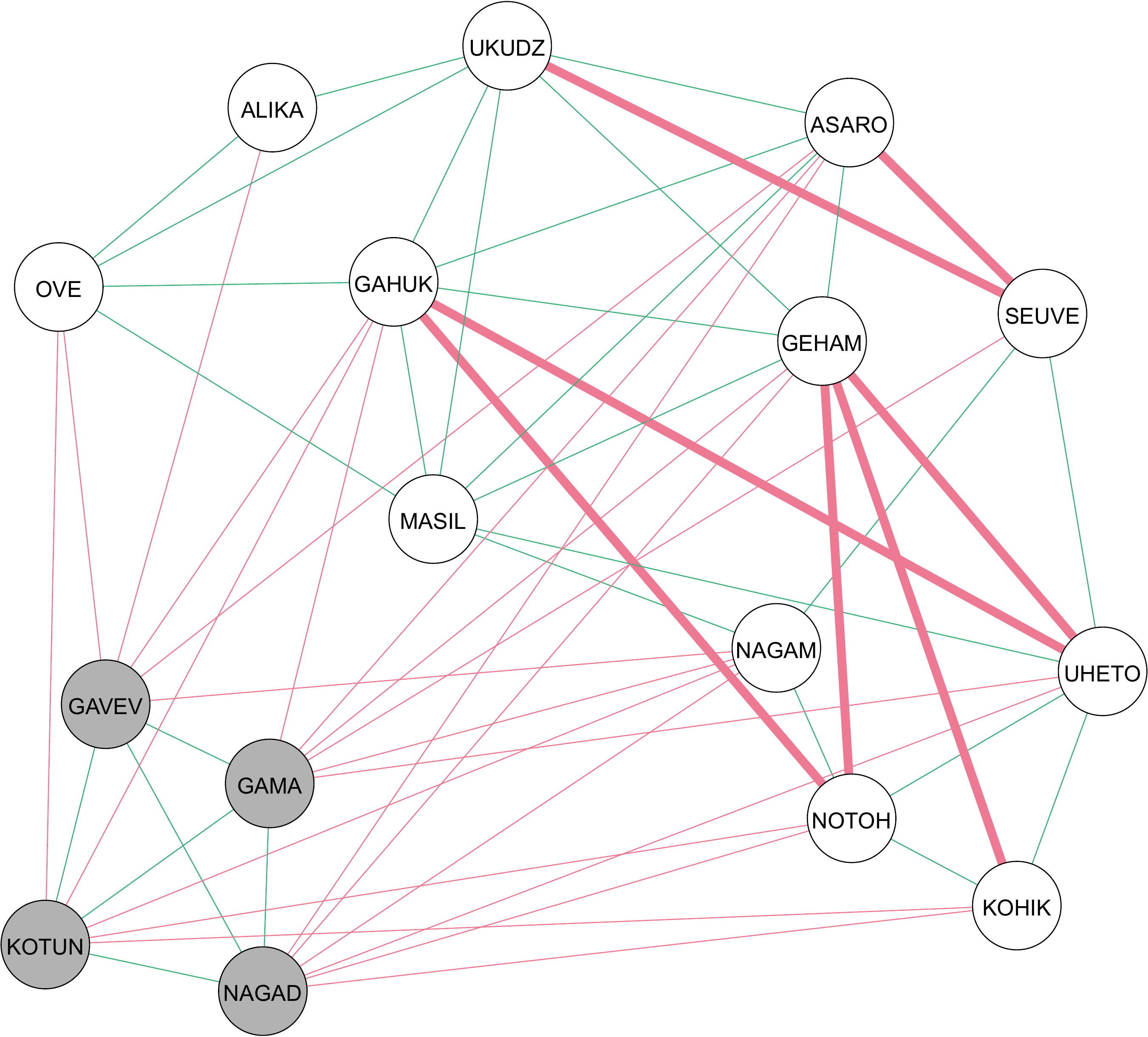}%
		\label{fig2a}}
	\hfil
	\centering
	\subfloat[Optimal colouring of monastery interactions network with frustrated edges shown by thick lines \cite{sampson_novitiate_1968}]{\includegraphics[height=1.5in]{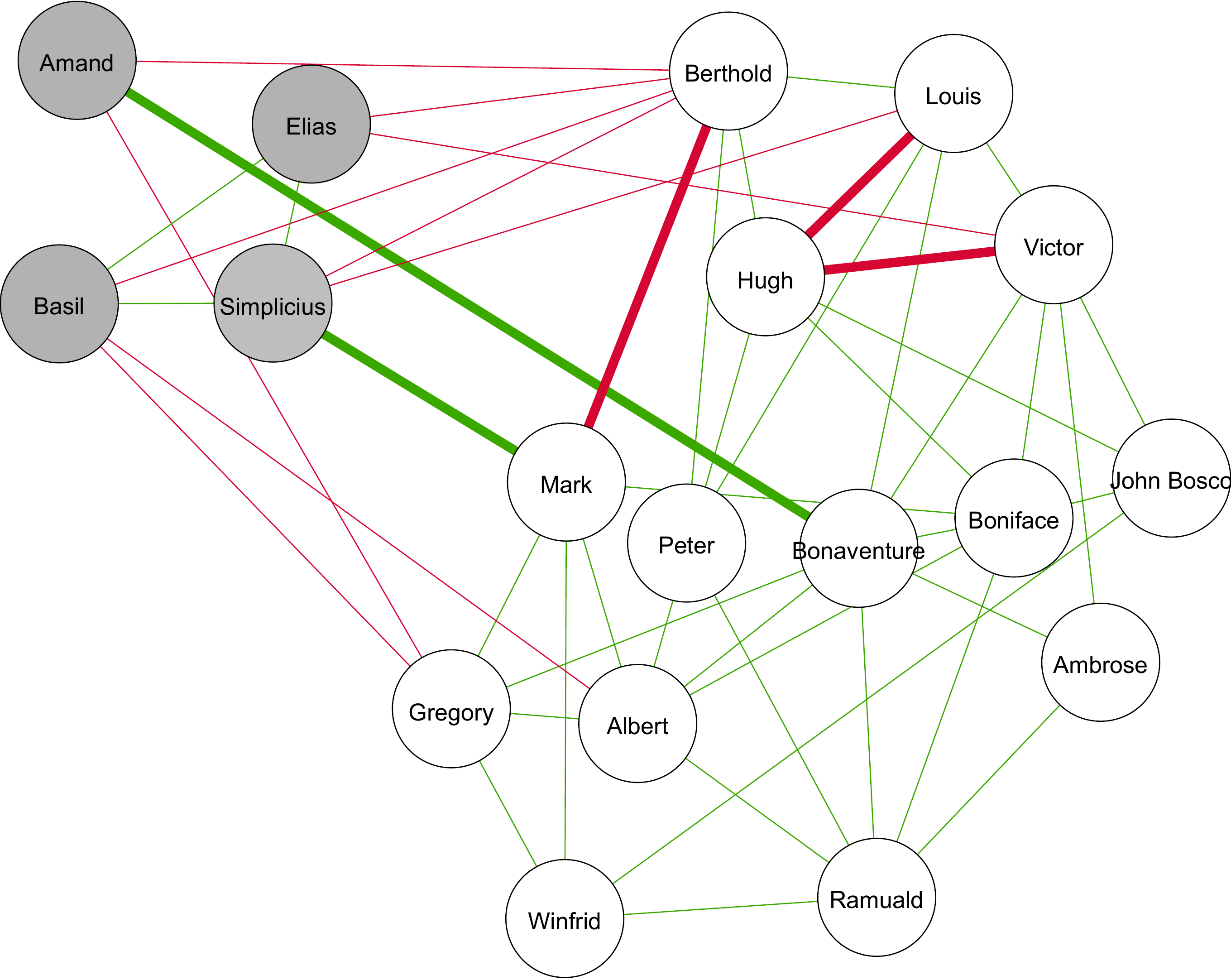}%
		\label{fig2b}} 
	\hfil
	\centering
	\subfloat[Network of Wikipedia elections \cite{leskovec_signed_2010}]{\includegraphics[height=1.7in]{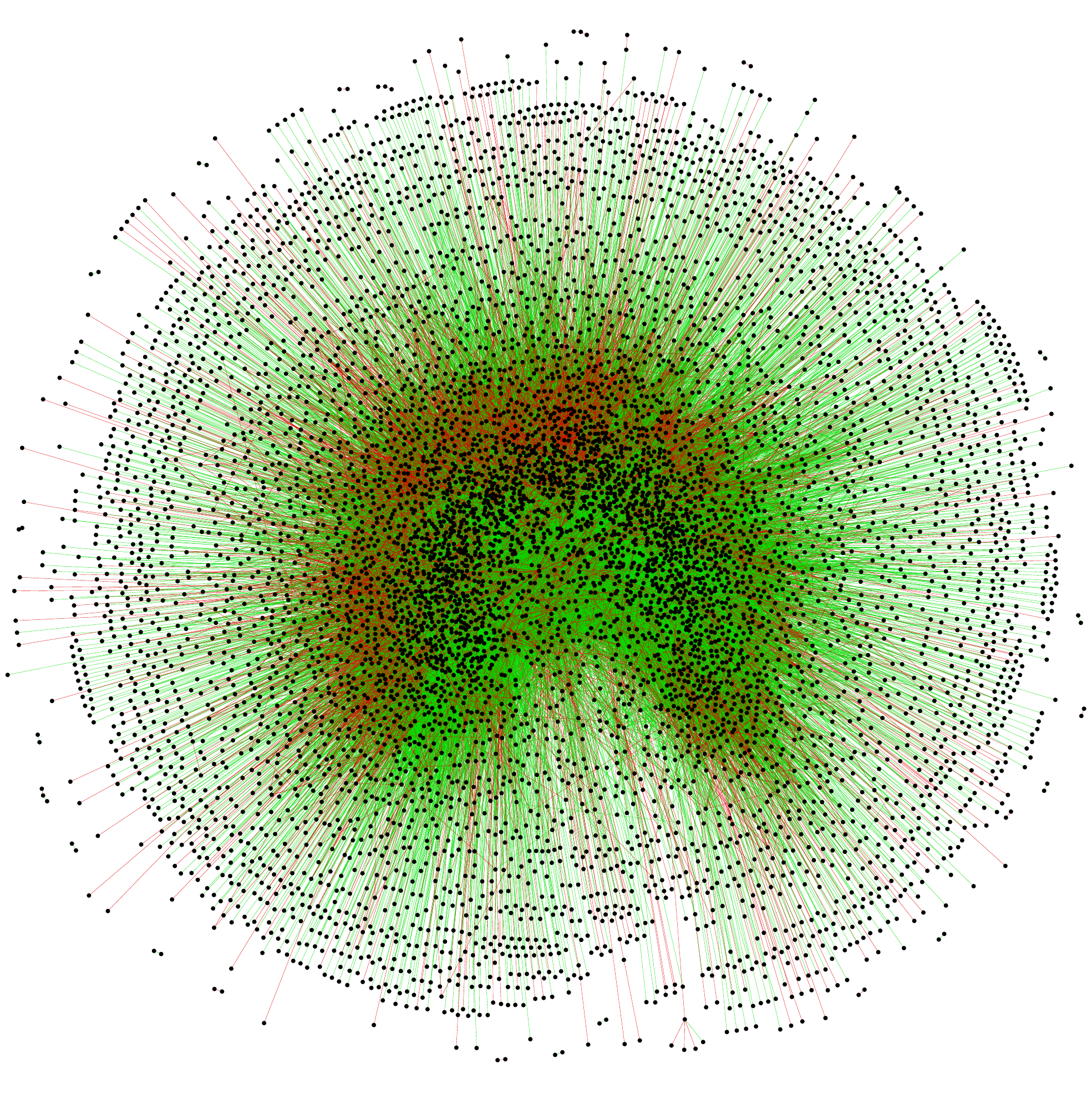}%
		\label{fig2d}} 
	\hfil
	\centering
	\subfloat[Network of the 108th US senate (nodes having mismatching party colour and optimal colour are positioned on the top and bottom.) \cite{neal_backbone_2014}]{\includegraphics[height=4in]{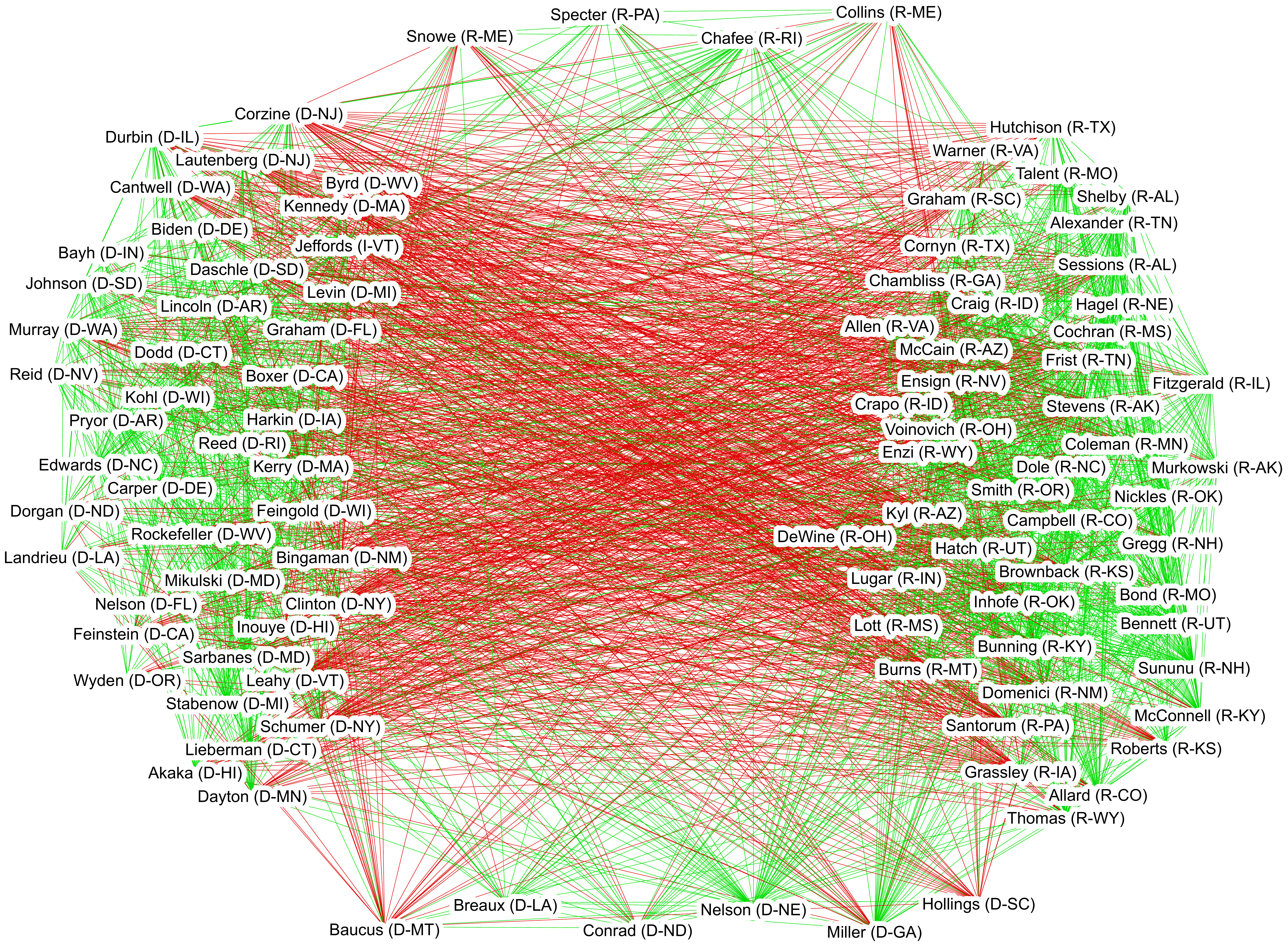}%
		\label{fig2c}} 
	\hfil
	\caption{Four signed networks inferred from the sociology and political science datasets and visualised using Gephi 
	}
	\label{fig2}
\end{figure}

A larger social signed network we use is from the Wikipedia election dataset \cite{leskovec_signed_2010}. This dataset is based on all adminship elections before January 2008 in which Wikipedia users have voted for approval or disapproval of other users promotions to becoming administrators. We use an undirected version of the Wikipedia elections signed graph made publicly available in \cite{levorato2015ils}. The four social signed networks are illustrated in Figure~\ref{fig2} where green and red edges represent positive and negative edges respectively. 

\subsection{Results}

Our numerical results are shown in Table~\ref{tab1} where the average and standard deviation of the frustration index in 500 reshuffled graphs (50 reshuffled graphs for Wikipedia election network), denoted by $L(G_r)$ and $\text{SD}$, are also provided for comparison. 

\FloatBarrier

\begin{table}[ht]
	\centering
	\caption{The frustration index in social signed networks}
	\label{tab1}
	\begin{tabularx}{\textwidth}{p{5.1cm}p{0.5cm}p{0.5cm}p{2.4cm}p{1.2cm}}
		\hline
		Graph $(n,m,m^-)$ & $\rho$ & $L(G)$ & $L(G_r) \pm \text{SD}$ & Z score \\ \hline
		Highland tribes $(16,58,29)$  & 0.483 & $7$    & $14.65 \pm 1.38$ &  $-5.54$ \\
		Monastery interactions $(18,49,12)$  & 0.320 & $5$    & $9.71 \pm 1.17$ &  $-4.03$ \\
		US senate $(100,2461,1047)$ & 0.497 & $331$  & $ 965.6\pm 9.08$& $-69.89$ \\ 
		Wiki elections $(7112,99917,21837)$ & 0.004 & $14532$&$18936.1 \pm 45.1 ^\dagger$  & $-97.59^\dagger$ \\
		\hline
		\multicolumn{4}{l}{$\dagger$ based on lower bounds within 15\% of the optimal solution}
	\end{tabularx}
\end{table}

As it is expected the four social signed networks are not totally balanced. However, the relatively small values of $L(G)$ suggest low levels of frustration in these networks. 
In order to be more precise, we have implemented a very basic statistical analysis using Z scores $Z={(L(G)-L(G_r))}/{\text{SD}}$. These Z scores, provided in the right column of the Table~\ref{tab1}, show how close the networks are to a state of balance. The results indicate that networks exhibit a level of frustration substantially lower than what is expected by chance.

The levels of balance in the three networks of Highland tribes, Monastery interactions, and US senate were previously measured in \cite{aref2015measuring} using not only the frustration index, but also measures based on triangles and eigenvalues \cite{kunegis_spectral_2010, terzi_spectral_2011, kunegis_applications_2014}. For these three networks which have comparatively high densities, the levels of balance measured using triangles and eigenvalues were reported to be consistent with the results that the frustration index provides \cite{aref2015measuring}.

\subsection{Partitions}

In the network of US senators, we may get insight not only from the value of the frustration index, but also the optimal node colouring leading to the minimum number of frustrated edges. As shown in node labels of Subfigure \ref{fig2c}, the 108th US senate was made of 48 Democratic senators, 1 independent senator (caucusing with Democrats), and 51 Republican senators. Based on these figures, we may consider a party colour for each node $i$ in the network and position senators from different parties on the left (Democrat) and right (Republican) sides of Subfigure \ref{fig2c}. The abundance of green edges on the left and right sides of Subfigure \ref{fig2c} shows that most senators have positive relationship with their party senators (co-sponsor bills proposed by their party). The numerous red edges between left and right sides of Subfigure \ref{fig2c} show opposition between senators from different parties (senators do not support bills put forward by the other party in most cases).

We can also compare the party colour for each node $i$ to the optimal colour ($x_i$ optimal value) that leads to the partitioning with the minimum number of frustrated edges. As expected from the bi-polar structure of US senate, the colours match in 90 out of 100 cases (considering the independent senator as a Democrat). The nodes associated with Republican (Democrat) senators who have mismatching party colour and optimal node colour are positioned on the top (bottom) of Subfigure \ref{fig2c}. It can be observed from such nodes that, contrary to the other nodes, they have negative (positive) ties to their (the other) party. 

\subsection{Computations}

Regarding performance of the optimisation model, a basic optimisation formulation of the problem with no speed-up technique (such as \cite[Eq.\ 8]{aref2017computing}) would solve the Highland tribes and Monastery interaction instances in a reasonable time on an ordinary computer \cite{aref2017computing}. The binary linear programming model \eqref{eq4} solves such instances in split seconds, while for the senators network it takes a few seconds.

For Wikipedia elections network, 9.3 hours of computation is required to find the optimal solution. This considerable computation time prevents us from testing 500 reshuffled versions of the Wikipedia elections network. In order to perform the statistical analysis for the Wikipedia network in a reasonable time, we limited the number of runs to 50.

As a conservative approach, we also used the average of best lower bound obtained within 15\% of the optimal solution as $L(G_r)$ for Wikipedia elections network. This approach reduces the average computation time of each run to 4 hours. The branch and bound algorithm guarantees that the frustration index of each reshuffled graph (which remains unknown) is greater than the lower bound we use to compute the Z score for Wikipedia elections network.

\section{Biological networks}\label{s:d2}

Some biological models are often used to describe interactions with dual nature between biological molecules in the field of systems biology. The interactions can be \textit{activation} or \textit{inhibition} and the biological molecules are enzymes, proteins or genes \cite{dasgupta_algorithmic_2007}. This explains the parallel between signed graphs and these types of biological networks. Interestingly, the concept of \textit{close-to-monotone} \cite{maayan_proximity_2008} in systems biology is analogous to being close to a state of balance. Similar to negative cycles and how they lead to unbalance, existence of \textit{negative loops} in biological networks indicates a system that does not display well-ordered behaviour \cite{maayan_proximity_2008}. 

\subsection{Datasets}

There are large-scale gene regulatory networks where nodes represent genes and positive and negative edges represent \textit{activating connections} and \textit{inhibiting connections} respectively. We use four signed biological networks previously analysed by \cite{iacono_determining_2010}. They include two gene regulatory networks, related to two organisms: a eukaryote (the yeast \textit{Saccharomyces cerevisiae}) \cite{Costanzo2001yeast} and a bacterium (\textit{Escherichia coli}) \cite{salgado2006ecoli}. Another signed network we use is based on the Epidermal Growth Factor Receptor (EGFR) pathway \cite{oda2005}. EGFR is related to the epidermal growth factor protein whose release leads to rapid cell division in the tissues where it is stored such as skin \cite{dasgupta_algorithmic_2007}. We also use a network based on the molecular interaction map of a white blood cell (macrophage) \cite{oda2004molecular}.

\begin{figure}[ht]
	\centering
	\subfloat[The gene regulatory network of the \textit{Escherichia coli} \cite{salgado2006ecoli}]{\includegraphics[height=1.7in]{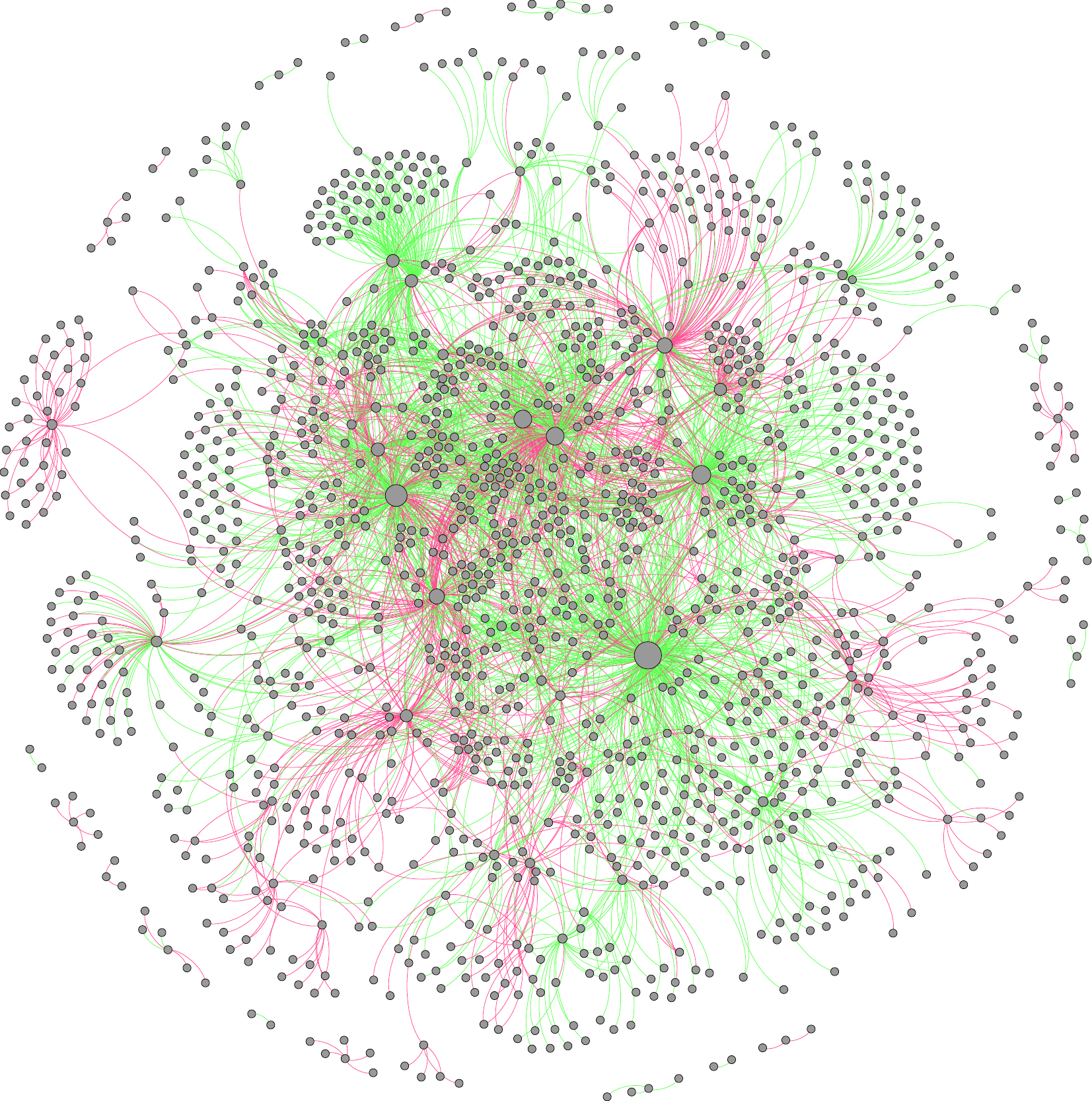}%
		\label{fig3b}} 
	\hfil
	\centering
	\subfloat[Epidermal growth factor receptor pathway \cite{oda2005}]{\includegraphics[height=1.6in]{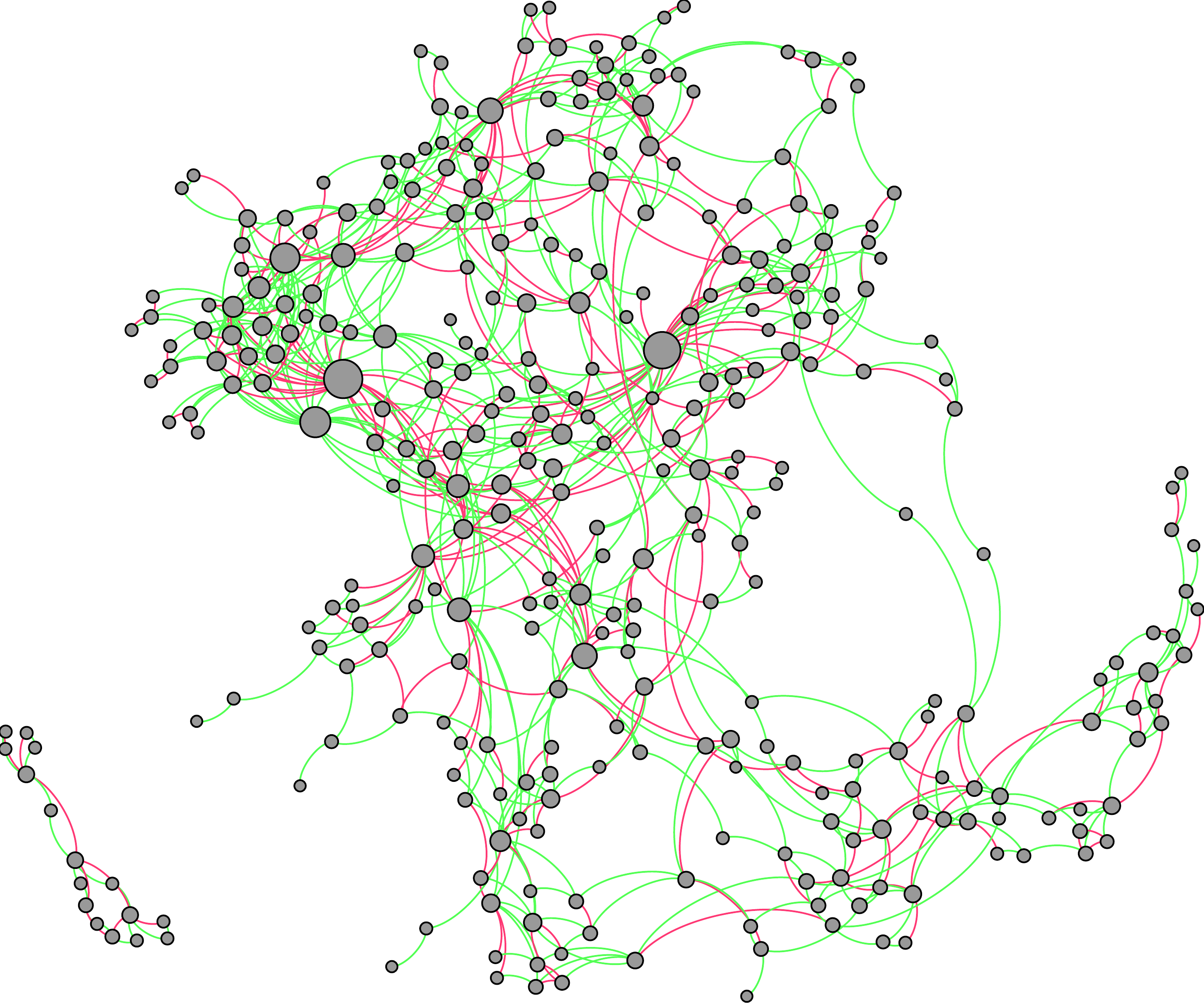}%
		\label{fig3c}} 
	\hfil
	\centering
	\subfloat[Molecular interaction map of a macrophage \cite{oda2004molecular}]{\includegraphics[height=1.6in]{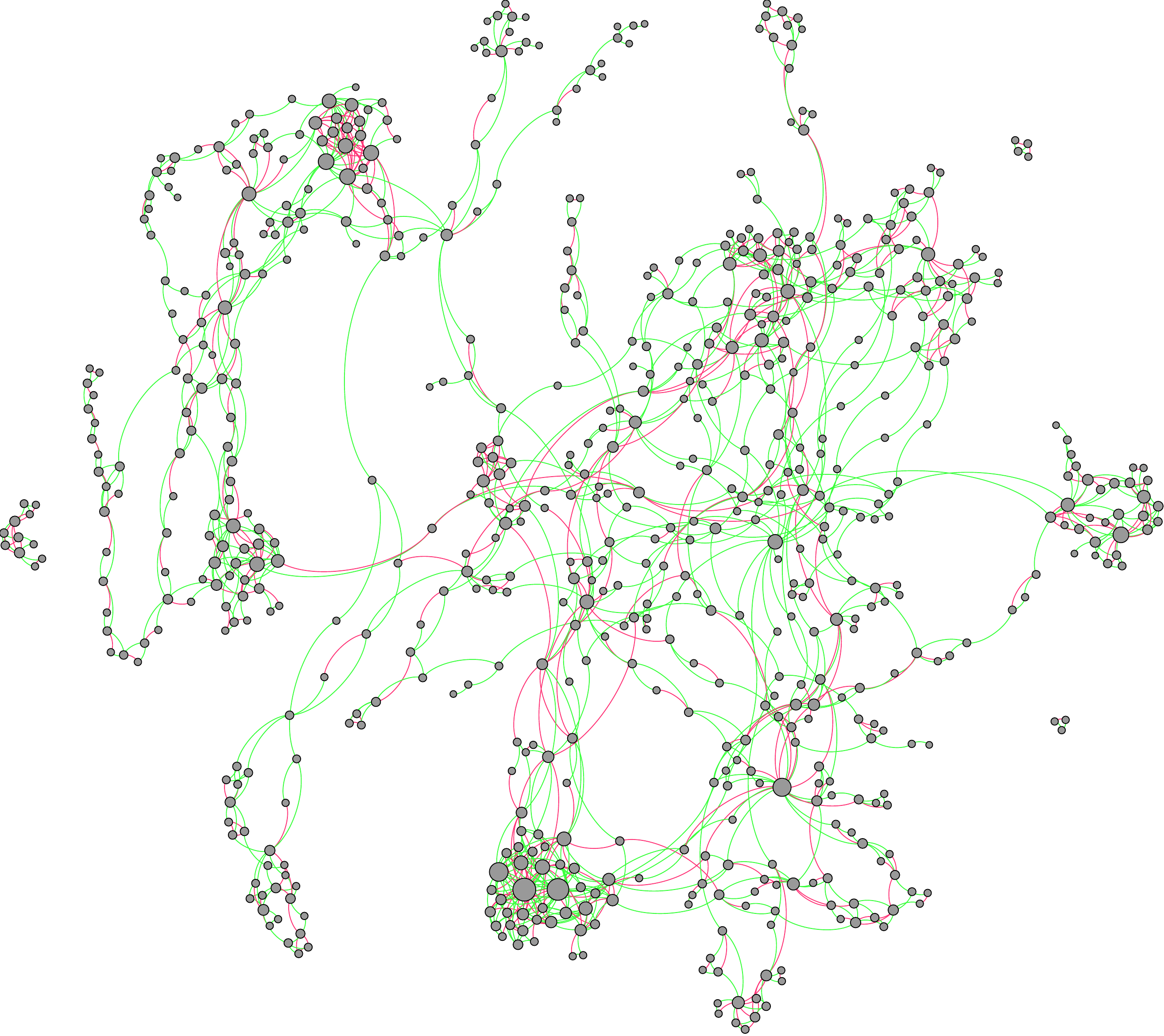}%
		\label{fig3d}} 
	\hfil
	\caption{Three biological signed networks visualised using Gephi 
}
	\label{fig3}
\end{figure}

Yeast and E.coli networks are categorised as transcriptional networks while EGFR and macrophage are signalling networks \cite{iacono_determining_2010}. Figure~\ref{fig3} shows three of these biological signed networks. The colour of edges correspond to the signs on the edges (green for activation and red for inhibition). For more details on the four biological datasets, one may refer to \cite{iacono_determining_2010}.

\subsection{Results}

Table \ref{tab2} provides the results for the four biological networks where the average and standard deviation of the frustration index in 500 reshuffled graphs are also provided for comparison.

\begin{table}[ht]
	\centering
	\caption{The frustration index in biological networks}
	\label{tab2}
	\begin{tabular}{lllll}
		\hline
		Graph $(n,m,m^-)$ & $\rho$ & $L(G)$ & $L(G_r) \pm \text{SD}$ & Z score \\ \hline
		yeast  $(690, 1080, 220)$ & 0.005 &  41  & $ 124.3\pm 4.97$& -16.75 \\ 
		E.coli  $(1461, 3215, 1336)$& 0.003 & 371  & $ 653.4\pm 7.71$& -36.64 \\ 	
		EGFR    $(329, 779, 264)$ & 0.014 & 193  & $ 148.96\pm 5.33$&   8.26 \\ 
		macrophage    $(678, 1425, 478)$ & 0.006 & 332  & $ 255.65\pm 8.51$&  8.98 \\  \hline

	\end{tabular}
\end{table}

The results in Table~\ref{tab2} show that the level of frustration is very low for yeast and E.coli networks. The Z score value for the yeast network is $-16.75$ which is consistent with the observation of DasGupta et al.\ (based on approximating the frustration index) that the number of edge deletions is up to 15 standard deviations away from the average for comparable random and reshuffled graphs \cite[Section 6.3]{dasgupta_algorithmic_2007}.

The Z score values in Table~\ref{tab2} show that the transcriptional networks are close to balanced (close-to-monotone) confirming observations in systems biology \cite{maayan_proximity_2008} that in such networks the number of edges whose removal eliminates negative cycles is small compared to the reshuffled networks. This explains the stability shown by such networks in response to external stimuli \cite{maayan_proximity_2008}.

\begin{figure}[ht]
	\centering
	\subfloat[The gene regulatory network of yeast \cite{Costanzo2001yeast}]{\includegraphics[height=2in]{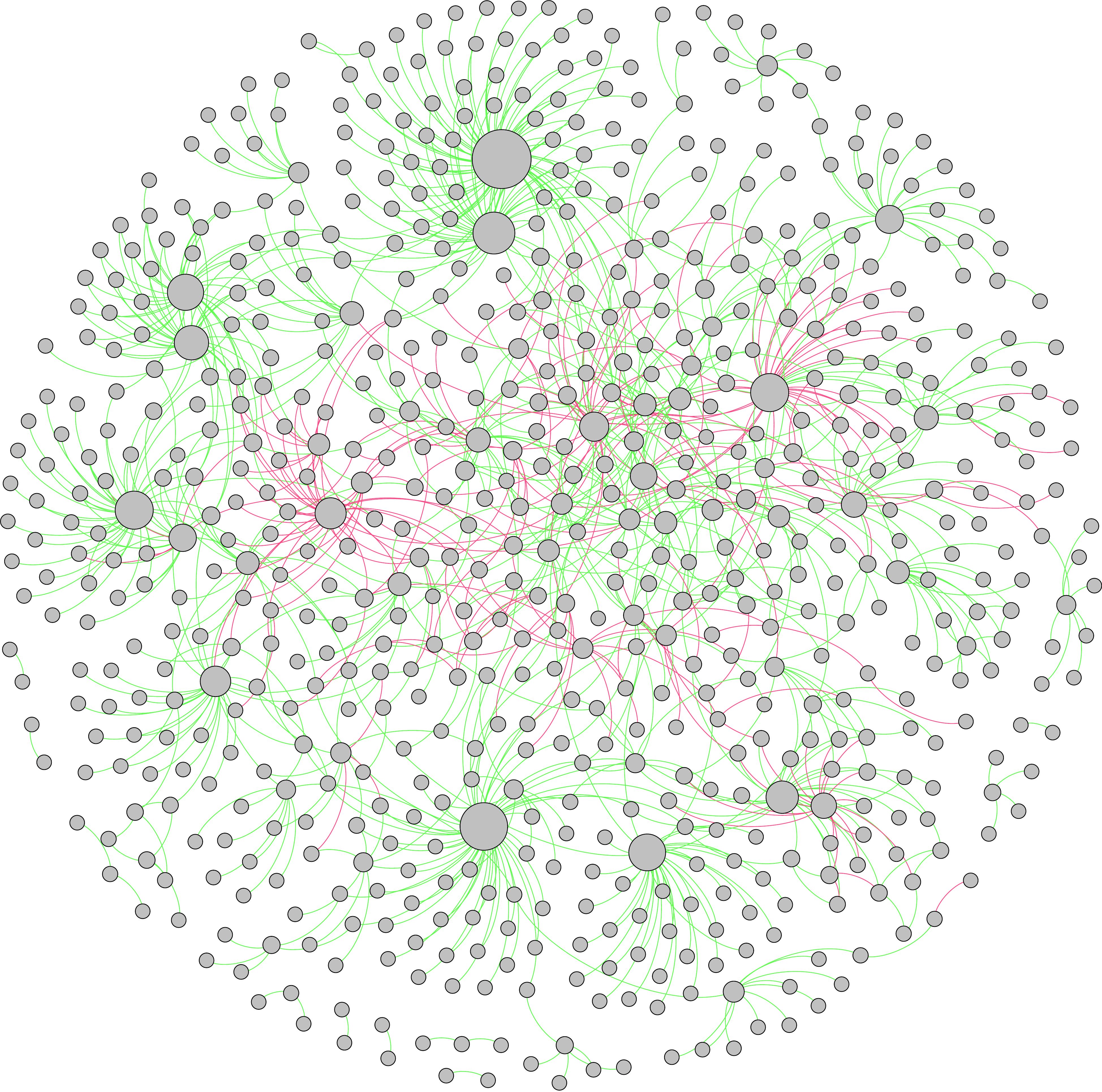}%
		\label{fig3a}} 
	\hfil
	\centering
	\subfloat[The monotone subsystem of yeast network]{\includegraphics[height=2in]{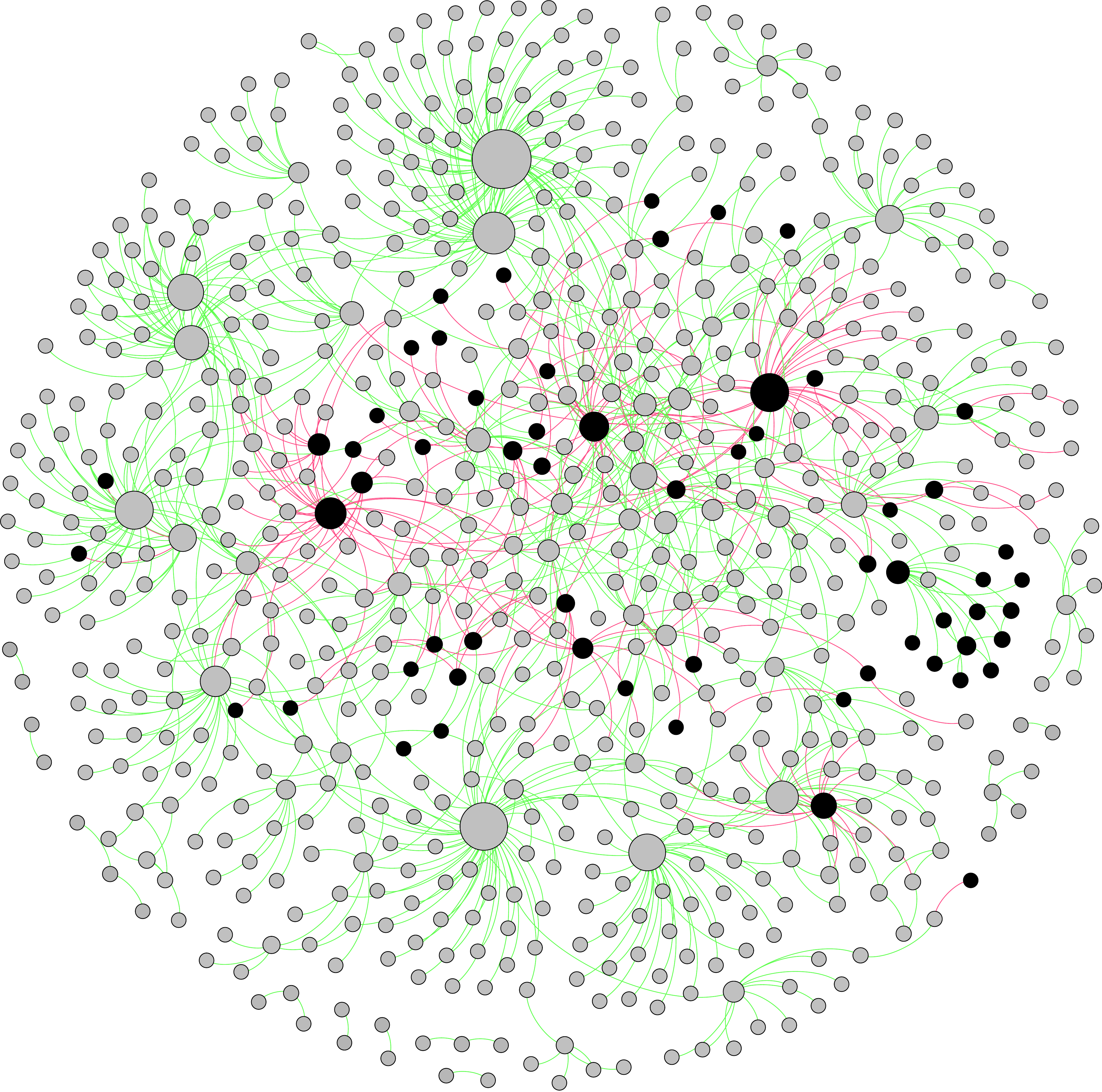}%
		\label{fig3e}} 
	\hfil
	\caption{The gene regulatory network (a) and monotone subsystem (b) of \emph{Saccharomyces cerevisiae} obtained after removing 41 frustrated edges and optimal node colours shown 
	}
	\label{fig3.5}
\end{figure}

Removing the frustrated edges from yeast network, we obtain the monotone subsystem (induced subgraph) whose response to perturbations can be predicted from the underlying structure \cite{dasgupta_algorithmic_2007, maayan_proximity_2008}. 
Figure \ref{fig3.5} shows the yeast network and its corresponding monotone subsystem obtained by removing the frustrated edges. In the monotone subsystem represented in \ref{fig3e} all walks connecting two given nodes have one specific sign. This prevents oscillation and chaotic behaviour \cite{dasgupta_algorithmic_2007, maayan_proximity_2008} and allows system biologists to predict how perturbing node A impacts on node B based on the sign of walks connecting A to B. All walks connecting nodes of the same colour (different colours) have a positive (negative) sign which leads to a monotone relationship between perturbation of one node and the impact on the other node.
In contrast for the two signalling networks, the level of frustration is very high, i.e., there are far more frustrated edges compared to the corresponding reshuffled networks. Networks of the EGFR protein and that of the macrophage are different from transcriptional networks in nature and our results show that they are far from balanced. This result is consistent with the discussions of Iacono et al.\ that EGFR and macrophage networks cannot be classified as close-to-monotone \cite[page 233]{iacono_determining_2010}.

Besides differences in network categories, one can see a structural difference between the transcriptional networks and signalling networks in Figures~\ref{fig3} -- \ref{fig3.5}. Subfigure \ref{fig3b} and Subfigure \ref{fig3a} show many high-degree nodes having mostly positive or mostly negative edges in the two transcriptional networks. However, such structures are not particularly common in the two signalling networks as visualised in Subfigures \ref{fig3c} -- \ref{fig3d}.

The levels of balance in the two networks of yeast and E.coli were previously measured in \cite{aref2015measuring} using the frustration index as well as measures based on triangles and eigenvalues \cite{kunegis_spectral_2010, terzi_spectral_2011, kunegis_applications_2014}. For these two networks, the levels of balance measured using eigenvalues and the frustration index were reported to be consistent. However, it was observed that the overall structure is not well reflected in the triangles because the small densities prevent the comparatively few triangles (70 triangles in yeast network and 1052 triangles in E.coli network) from representing the closeness of the two networks to balance \cite{aref2015measuring}.

\subsection{Computations}

The two smallest biological networks considered here (EGFR and macrophage) are the largest networks analysed in a recent study of balancing signed networks by negating minimal edges \cite{Wang2016} in which the heuristic algorithm gives sub-optimal values of the frustration index \cite[Fig. 5]{Wang2016}. 

Aref et al.\ compare the quality and solve time of their exact algorithm with that of recent heuristics and approximations implemented on the same datasets  \cite{aref2016exact,aref2017computing}. While data reduction schemes \cite{huffner_separator-based_2010} may take up to 1 day for these four biological networks and heuristic algorithms \cite{iacono_determining_2010} only provide bounds with up to 9\% gap from optimality, the binary linear programming model \eqref{eq4} equipped with the speed-up techniques reaches global optimality in a few seconds on an ordinary computer \cite{aref2016exact,aref2017computing}. 

\section{International relations} \label{s:temporal}\label{s:d3}

International relations between countries can be analysed using signed networks models and balance theory \cite{lai_alignment_2001, doreian_partitioning_2013, Lerner2016}. In earlier studies of balance theory, Harary used signed relations between countries over different times as an example of balance theory applications in this field \cite{harary1961structural}.

\subsection{Datasets}

In this section, we analyse the frustration index in a temporal political network of international relations. Doreian and Mrvar have used the Correlates of War (CoW) datasets \cite{correlatesofwar2004} to construct a signed network with 51 sliding time windows each having a length of 4 years \cite{patrick_doreian_structural_2015}. Joint memberships in alliances, being in unions of states and sharing inter-governmental agreements are represented by positive edges. Being at war (or in conflict without military involvement) and having border disputes or sharp disagreements in ideology or policy are represented by negative edges \cite{patrick_doreian_structural_2015}.

A dynamic visualisation of the network can be viewed on the \urllink{https://youtu.be/STlNsTjYjAQ}{YouTube video sharing website} \cite{aref_youtube}. This temporal network represents more than half a century of international relations among countries in the post Second World War era starting with 1946-1949 time window and ending with 1996-1999 time window \cite{patrick_doreian_structural_2015}. One may refer to \cite[Section 3.4]{patrick_doreian_structural_2015} for a detailed explanation of using sliding time windows and other details involved in constructing the network. In the first time window of the temporal network, network parameters are $n=64$, $m=362$ and $m^-=42$. 
In the last time window, these parameters are $n=151$, $m=1247$ and $m^-=147$. 

\subsection{Results}
\begin{figure}[ht]
	\centering
	\includegraphics[width=0.8\textwidth]{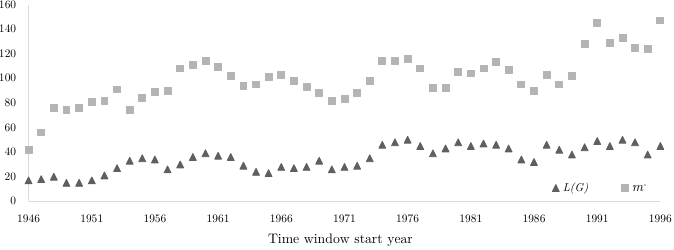}
	\caption{The number of negative edges $m^-$ and the frustration index $L(G)$ of the CoW dataset over time}
	\label{fig4}
\end{figure}

Figure \ref{fig4} demonstrates the number of negative edges and the frustration index in the CoW dataset. Doreian and Mrvar have attempted analysing the signed international network using the frustration index (under a different name) \cite{patrick_doreian_structural_2015} and other measures. They used a blockmodeling algorithm in \textit{Pajek} for obtaining the frustration index. However, their solutions are not optimal and thus do not give the frustration index for any of the 51 instances.

Even with reliable numerical results in hand, caution must be applied before answering whether this network has become closer to balance over the time period 1946-1999 \cite{harary_signed_2002, antal_social_2006, marvel_continuous-time_2011} or the simpler question, how close this network is to total balance \cite{aref2015measuring}. 

\begin{figure}[ht]
	\centering
	\includegraphics[width=0.75\textwidth]{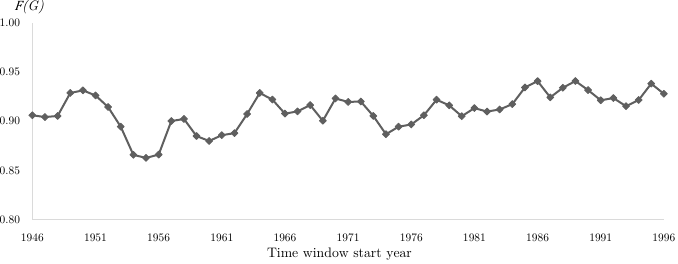}
	\caption{The normalised frustration index in the CoW dataset over time}
	\label{fig5}
\end{figure}

Using Z scores, we observe tens of standard deviation difference between the frustration index of 51 CoW instances and the average frustration index of the corresponding reshuffled graphs. This indicates that the network has been comparatively close to a state of structural balance over the 1946-1999 period. This is contrary to the evaluation of balance by Doreian and Mrvar using their frustration index estimates \cite{patrick_doreian_structural_2015}. Bearing in mind that the size and order changes in each time window of the temporal network, we use the normalised frustration index, $F(G)=1-2L(G)/m$, in order to investigate the partial balance over time. $F(G)$ provides values in the unit interval where the value $1$ represents total balance. Figure~\ref{fig5} shows that all normalised frustration index values are greater than $0.86$. 

The data plotted in Figure~\ref{fig5} can also be used to statistically test the stationarity of the normalised frustration index values. The Priestley-Subba Rao (PSR) test of non-stationarity \cite{priestley_test_1969} provides the means of a statistically rigorous hypothesis testing for stationarity of time series. We use an R implementation of the PSR test that is available in the \textit{fractal} package in the CRAN repository. The p-value of non-stationarity test for variation of $F(G)$ over time equals $0.03$ indicating that there is strong evidence to reject the null hypothesis of stationarity.

While there is no monotone trend in the values of $F(G)$, in most years the network has moved towards becoming more balanced over the 1946-1999 period. The overlap of the time period with the Cold War era may explain how the network has been close to a global state of bi-polarity with countries clustered into two antagonist sides. Doreian and Mrvar claim to have decisive evidence \cite{patrick_doreian_structural_2015} (based on frustration index estimates not showing monotonicity) against the theory \cite{harary_signed_2002, antal_social_2006, marvel_continuous-time_2011} that signed networks evolve towards becoming more balanced. Our observations based on $F(G)$ values do not reject this theory.

\subsection{Partitions}
We can investigate how the 180 countries of the CoW dataset are partitioned into two internally solidary but mutually hostile groups in this network. The optimal node colours show that 32 countries have remained in one fixed partition, which we call group A, over the 1946-1999 period. Group A includes Argentina, Belgium, Bolivia, Brazil, Canada, Chile, Colombia, Costa Rica, Denmark, Dominican Republic, Ecuador, El Salvador, France, Great Britain, Guatemala, Haiti, Honduras, Iceland, Italy, Luxembourg, Mexico, Netherlands, Nicaragua, Norway, Panama, Paraguay, Peru, Portugal, Turkey, Uruguay, United States, and Venezuela. There are also 26 other countries that mostly (in over 40 time frames) belong to same partition as group A.
\begin{figure}[ht]
	\centering
	\includegraphics[width=0.9\textwidth]{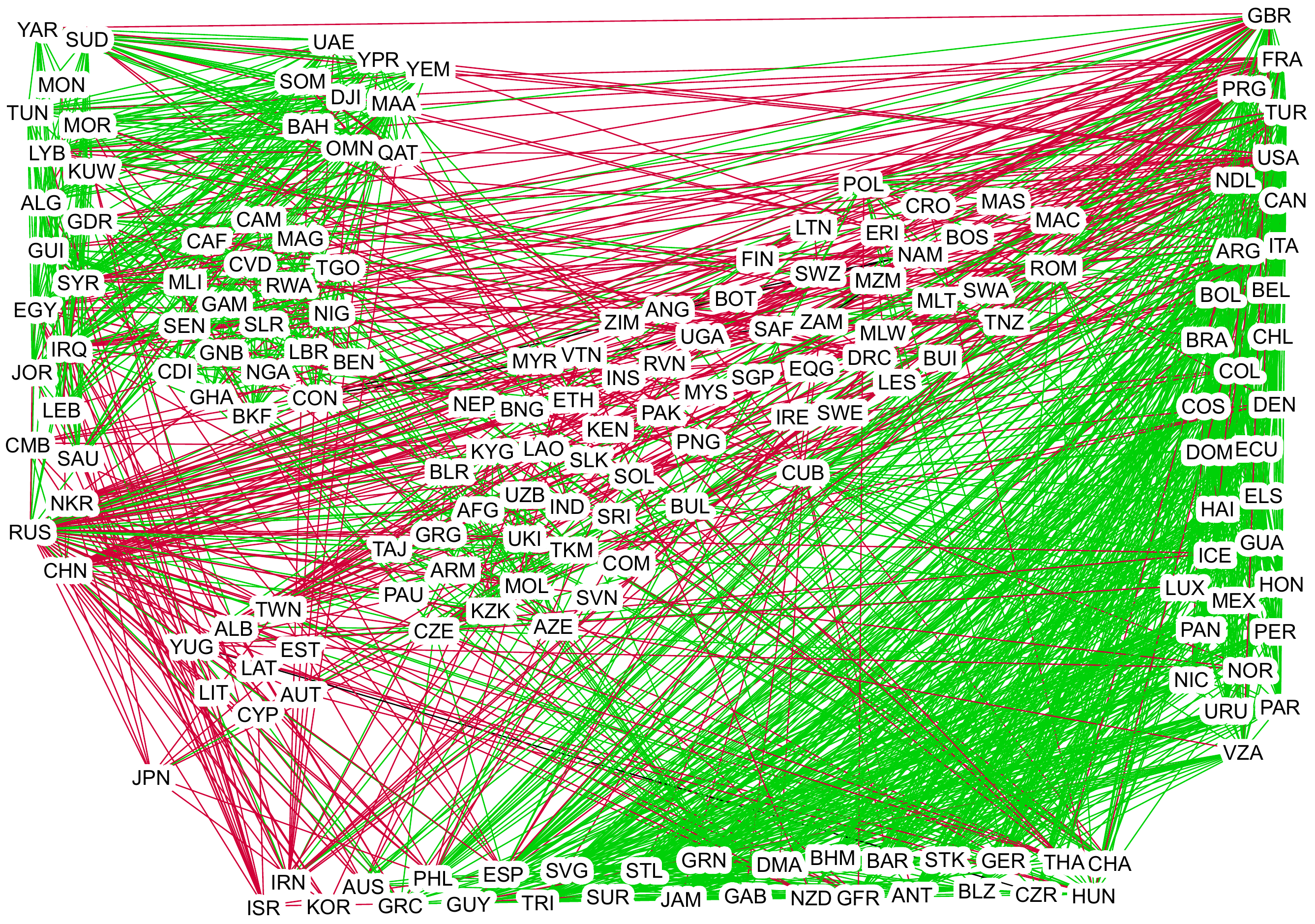}
	\caption{The partitioning of countries into groups A (right) and B (left) with most intra-group (inter-group) edges being positive (negative), the countries positioned in the bottom mostly belong to the same partition as group A.}
	\label{fig5.5}
\end{figure}

20 countries, which we call group B, oppose group A in over 40 time frames. Group B includes North Korea, Sudan, Tunisia, Morocco, Libya, Kuwait, Algeria, German Democratic Republic (East Germany), Guinea, Syria, Egypt, Iraq, Jordan, Lebanon, Russia, Saudi Arabia, Cambodia, Mongolia, China, and Yemen.

The remaining 102 countries change partition several times over the 1946-1999 period. Figure~\ref{fig5.5} shows the partitioning of the countries in which groups A and B are positioned on the right and left sides respectively. The countries that mostly belong to group A partition are positioned at the bottom of Figure~\ref{fig5.5}. The countries more inclined towards group B are positioned closer to the top left side of Figure~\ref{fig5.5}.

\subsection{Computations}
For this dataset, the binary linear programming model provides the exact values of the frustration index in less than $0.1$ seconds on an ordinary computer with an Intel Core i5 7600 @ 3.50 GHz processor and 8.00 GB of RAM \cite{aref2016exact}.

\section{Financial portfolios}\label{ss:finance}\label{s:d4}

There are studies investigating financial networks of securities modelled by signed graphs \cite{harary_signed_2002,huffner_separator-based_2010, figueiredo2014maximum}. Harary et al.\ originally suggested analysing portfolios using structural balance theory \cite{harary_signed_2002}. They represented securities of a portfolio by nodes and the correlations between pairs of securities by signed edges \cite{harary_signed_2002}. They used $\pm0.2$ as thresholds for considering a signed edge between two securities of a portfolio. Simplifying a portfolio containing Dow Jones, London FTSE, German DAX, and Singapore STI to a signed graph with four nodes, they observed that the graph has remained in a state of balance from October 1995 to December 2000 \cite{harary_signed_2002}. H\"{u}ffner, Betzler, and Niedermeier considered portfolios containing 60-480 stocks and thresholds of $\pm0.325,\pm0.35,\pm0.375$ to evaluate the scalability of their algorithm for approximating the frustration index \cite{huffner_separator-based_2010}. Their dataset is also analysed in \cite{figueiredo2014maximum}.

\subsection{Datasets}
In this subsection, we consider well-known portfolios recommended by financial experts for having a low risk in most market conditions \cite{bogle2015}. These portfolios are known as lazy portfolios and usually contain a small number of \textit{well-diversified} securities \cite{bogle2015}. We consider 6 lazy portfolios each consisting of 5-11 securities. Table~\ref{tab3} represents the six lazy portfolios and their securities.

\begin{figure}[ht]
	\subfloat[A weighted complete graph representing correlation coefficients, edge weights $\leq -0.2$ are shown on the edges]{\includegraphics[height=2.2in]{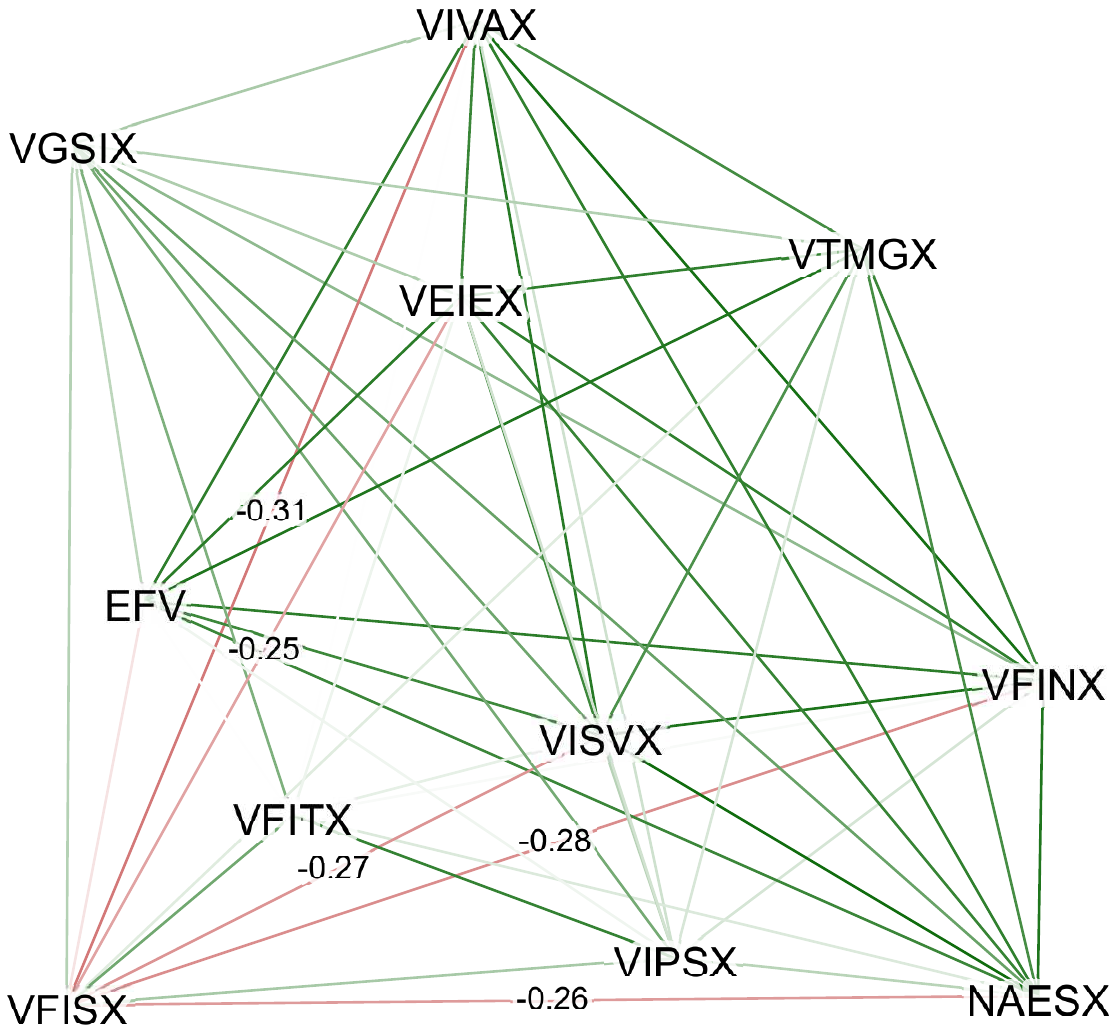}%
		\label{fig6a}} 
	\hfil
	\subfloat[The portfolio signed graph produced by thresholding on $\pm0.2$, frustrated edges are shown by thick lines]{\includegraphics[height=2.2in]{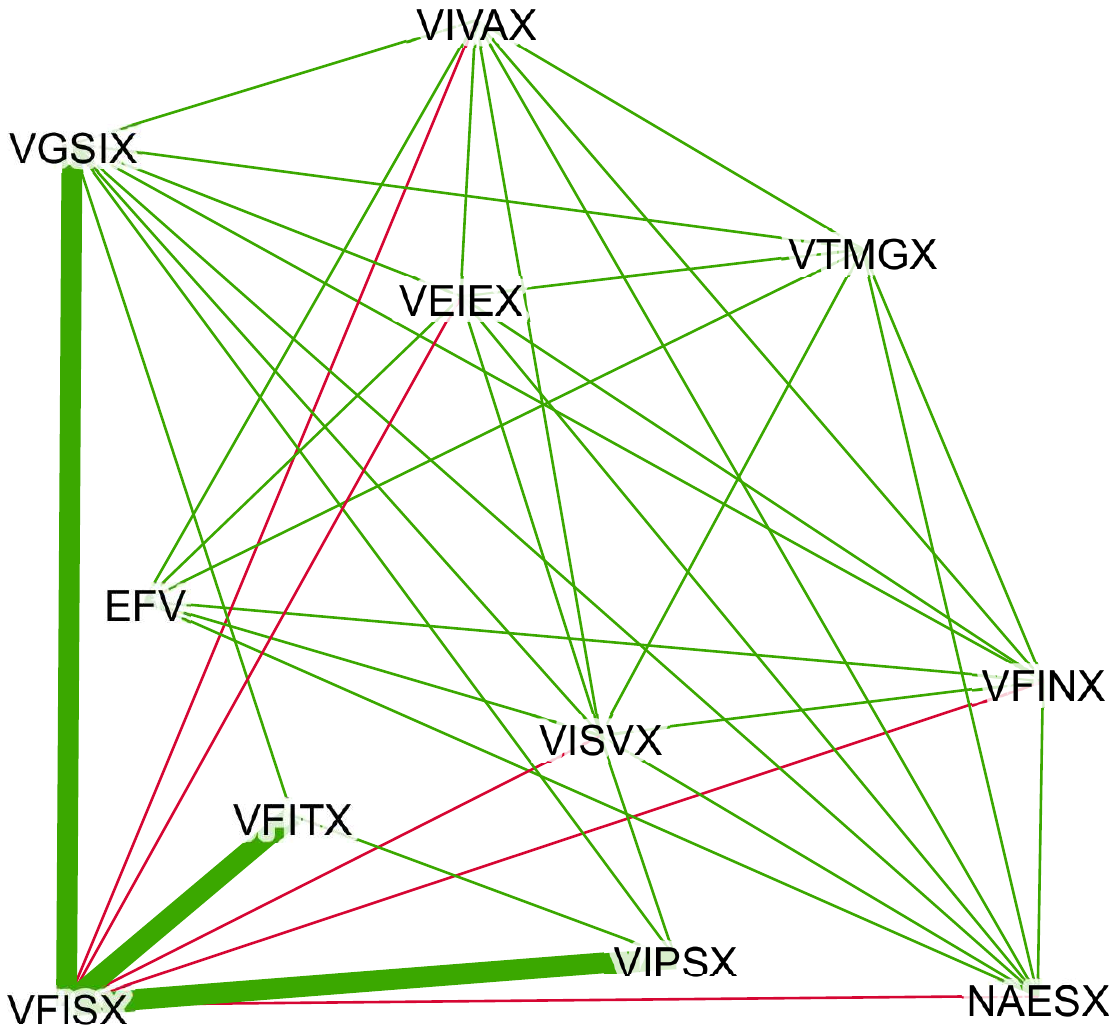}%
		\label{fig6b}} 
	\hfil
	\caption{Portfolio P3 in 2016-10 (unbalanced) illustrated as (a) weighted and (b) signed networks using Gephi 
	}
	\label{fig6}
\end{figure}

\begin{table}[ht]
	\centering
	\caption{Six portfolios and their securities}
	\label{tab3}
	\begin{tabular}{p{1cm}p{1cm}p{1cm}p{1cm}p{1cm}p{1.3cm}p{1cm}}
		\hline
		Portfolio & \urllink{http://tinyurl.com/y88fkv67}{Ivy portfolio (P1)} & \urllink{http://tinyurl.com/yc5scb9z}{Simple portfolio (P2)} & \urllink{http://tinyurl.com/y74k72e9}{Ultimate Buy \& Hold (P3)} & \urllink{http://tinyurl.com/y922s5f2}{Yale Endowment (P4)} & \urllink{http://tinyurl.com/ybtx7295}{Swensen's lazy portfolio (P5)} & \urllink{http://tinyurl.com/ybzalvzz}{Coffee House (P6)} \\ \hline
		Financial Expert&Mebane Faber&Larry Swedroe&Paul Merriman&David Swensen&David Swensen&Bill Schultheis\\ \hline
		VEIEX                                                                          &    & x  & x  & x  & x  &                        \\
		VGSIX                                                                          &    &    & x  & x  & x  & x                      \\
		VIPSX                                                                          &    & x  & x  & x  & x  &                        \\
		VTMGX                                                                          &    &    & x  & x  & x  &                        \\
		VIVAX                                                                          &    & x  & x  &    &    & x                      \\
		NAESX                                                                          &    & x  & x  &    &    & x                      \\
		EFV                                                                            &    & x  & x  &    &    &                        \\
		VFINX                                                                          &    &    & x  &    &    & x                      \\
		VFISX                                                                          &    &    & x  &    & x  &                        \\
		VISVX                                                                          &    &    & x  &    &    & x                      \\
		VTSMX                                                                          &    &    &    & x  & x  &                        \\
		IJS                                                                            &    & x  &    &    &    &                        \\
		TLT                                                                            &    &    &    & x  &    &                        \\
		VFITX                                                                          &    &    & x  &    &    &                        \\
		VBMFX                                                                          &    &    &    &    &    & x                      \\
		VGTSX                                                                          &    &    &    &    &    & x                      \\
		GSG                                                                            & x  &    &    &    &    &                        \\
		IEF                                                                            & x  &    &    &    &    &                        \\
		VEU                                                                            & x  &    &    &    &    &                        \\
		VNQ                                                                            & x  &    &    &    &    &                        \\
		VTI                                                                            & x  &    &    &    &    &                        \\ \hline
		$n$		&5 &6 &11 &6 &6 &7\\ \hline
	\end{tabular}
\end{table}
\FloatBarrier
The signed networks representing the lazy portfolios are generated by considering prespecified thresholds as in \cite{harary_signed_2002,huffner_separator-based_2010}. We use the daily returns correlation data that can be found on the Portfolio Visualizer website \cite{Portfolio_visualizer} and thresholds of $\pm0.2$ similar to \cite{harary_signed_2002}. Correlation coefficients with an absolute value greater than $0.2$ are considered to draw signed edges between the securities with respect to the sign of correlation. Figure \ref{fig6} shows two networks of portfolio P3 based on the October 2016 data. The nodes represent 11 securities of the portfolio and the colours of the edges correspond to the correlations between the securities (green for positive and red for negative correlation). Lighter colours in Figure~\ref{fig6} (a) represent smaller absolute values of correlation coefficient.

\subsection{Results}
\begin{figure}[ht]
	\centering
	\includegraphics[width=1\textwidth]{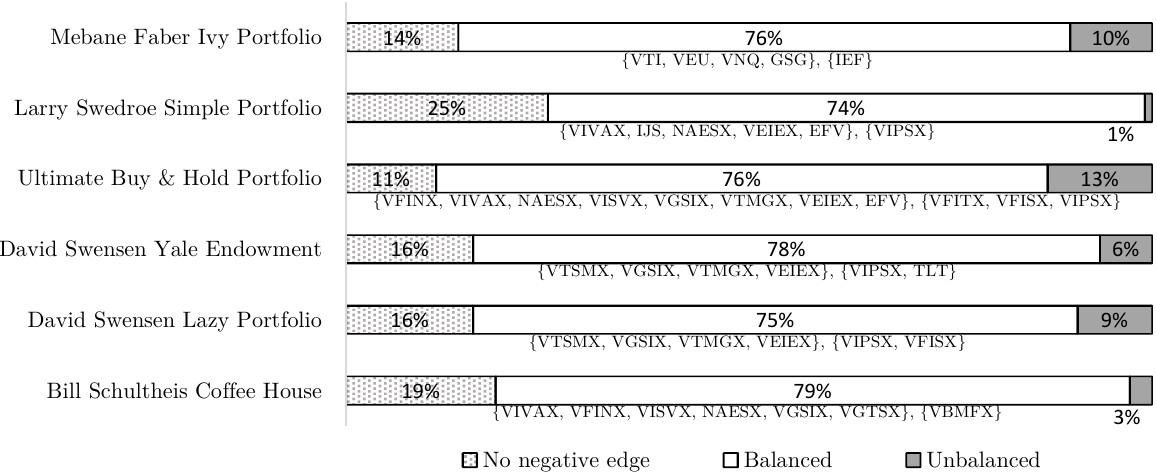}
	\caption{Frequencies of all-positive, balanced, and unbalanced networks over 108 monthly time frames}
	\label{fig7}
\end{figure}

We analyse 108 monthly time frames for each of the six portfolios which correspond to the signed networks of each month within the 2008-2016 period. 
The signed networks obtained are totally balanced and have negative edges in a large number of time frames (74-79\%). In a relatively small number of time frames (1-13\%), the underlying network is unbalanced. Figure~\ref{fig7} illustrates the results which are consistent with the findings of Harary et al.\ in \cite{harary_signed_2002} in terms of balanced states being dominant. 
\FloatBarrier

\begin{figure}[ht]
	\centering
	\includegraphics[width=1\textwidth]{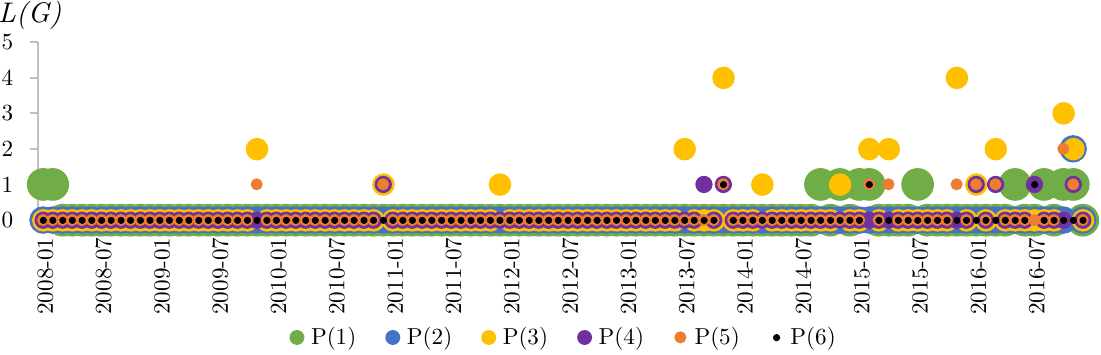}
	\caption{Frustration index of six portfolios over 108 monthly time frames 
	}
	\label{fig9}
\end{figure}

More detailed results on balance states and frustration index of six portfolios over time are provided in Figure~\ref{fig9}. We observe that there are some months when several portfolios have an unbalanced underlying signed graph (non-zero frustration index values). One may suggest that common securities explain this observation, but P(1) does not have any security in common with other portfolios which suggests otherwise. It can be observed from Figure~\ref{fig9} that non-zero frustration index values are rather rare and usually very small.

\subsection{Partitions}
The optimal partitioning of each portfolio into two sub-portfolios (with positive correlations within and negative correlations in between) can be obtained from optimal node colours. The optimal partitioning remains mostly unchanged over balanced states among 108 time frames. Figure~\ref{fig7} also shows the most common optimal partitioning of the securities for each portfolio. 

\subsection{Computations}
Regarding sensitivity of the results to the cut-off threshold, other thresholds (like $\pm0.1$ and $\pm0.3$) also lead to balanced states being dominant. Using thresholds of $\pm0.1$ leads to relatively more unbalanced states and less all-positive states, while thresholds of $\pm0.3$ have the opposite effect. Regarding computational performance for these small instances, a basic optimisation formulation of the problem with no speed-up technique (such as \cite[Eq.\ 8]{aref2017computing}) would solve the portfolio instances in a reasonable time on an ordinary computer.

\FloatBarrier
\section{Closely related problems from chemistry and physics}\label{s:d56}

In this section, we briefly discuss two problems from chemistry and physics that are closely related to the frustration index of signed graphs. The parallels between these problems and signed graphs allow us to use the binary linear model in \eqref{eq4} to tackle the NP-hard computation of important measures for relatively large instances. We discuss computation of a chemical stability indicator for carbon molecules in Subsection \ref{s:d5} and the optimal Hamiltonian of Ising models in Subsection \ref{s:d6}.

\subsection{Bipartivity of fullerene graphs} \label{ss:chemistry}\label{s:d5}
\begin{figure}[ht]
	\subfloat[Fullerene graph of C240]{\includegraphics[height=1.2in]{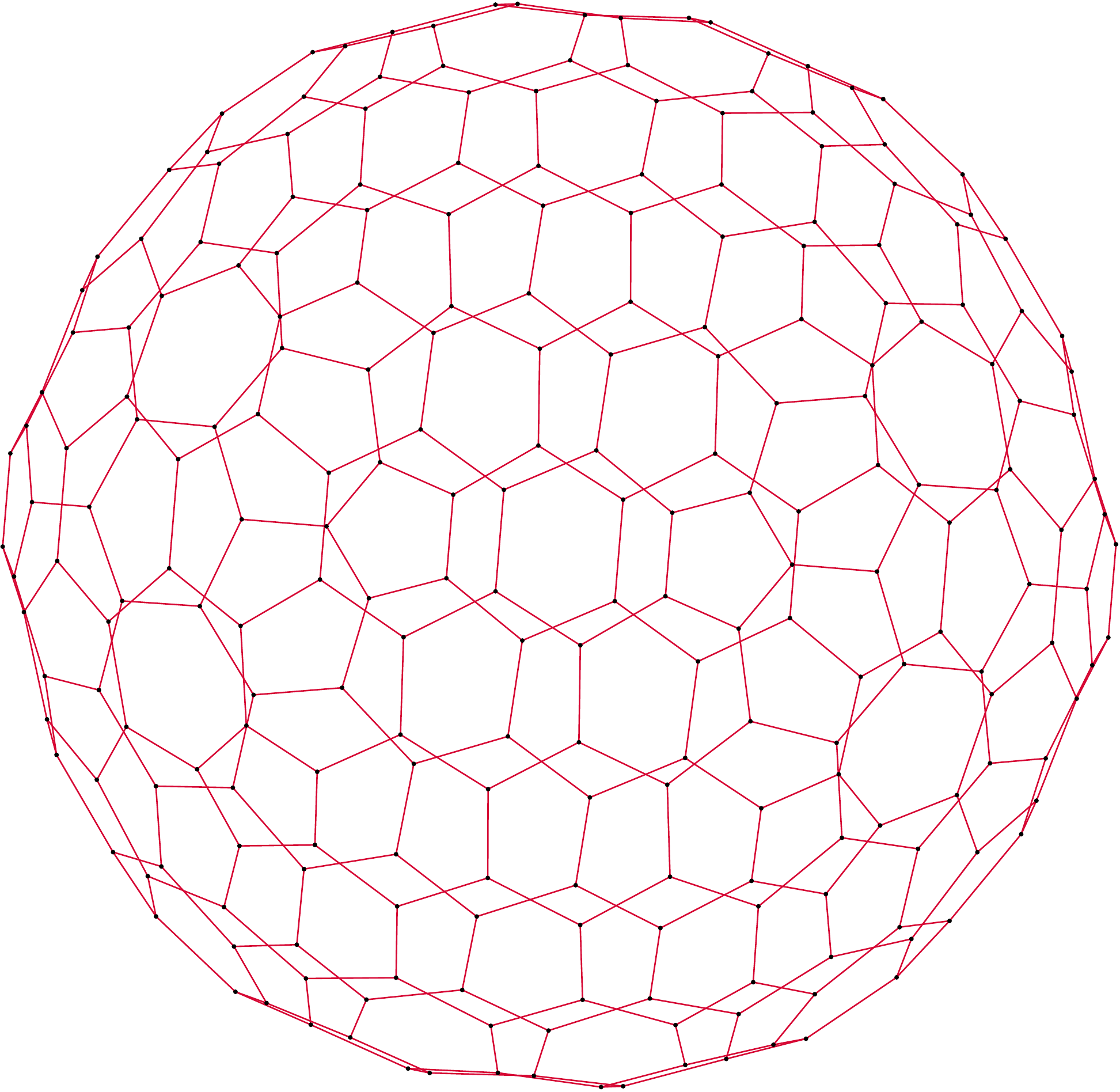}%
		\label{fig10a}} 
	\hfil
	\subfloat[Fullerene graph of C240 made bipartite after removing 24 edges]{\includegraphics[height=1.2in]{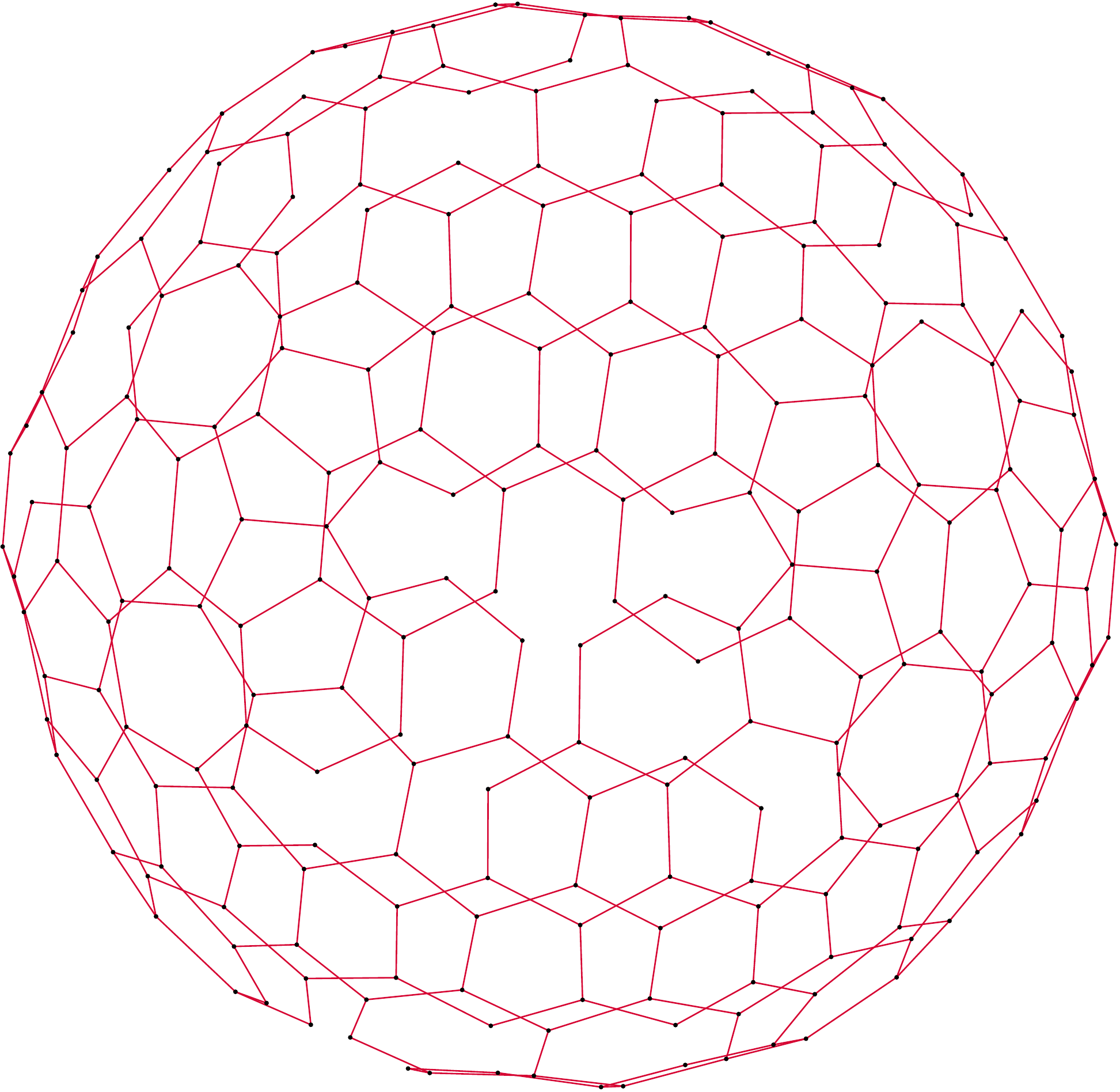}%
		\label{fig10b}} 
	\hfil
	\subfloat[Fullerene graph of C2160]{\includegraphics[height=1.2in]{fig11c-eps-converted-to.pdf}%
		\label{fig10c}} 
	\hfil
	\subfloat[Fullerene graph of C6000]{\includegraphics[height=1.2in]{fig11d-eps-converted-to.pdf}%
		\label{fig10d}} 
	\caption{Several fullerene graphs represented as signed graphs in which all edges are declared to be negative 
	}
	\label{fig10}
\end{figure}

Previous studies by Do{\v{s}}li{\'c} and associates suggest that graph bipartivity measures are potential indicators of chemical stability for carbon structures known as fullerenes \cite{doslic2005bipartivity, doslic_computing_2007}. The graphs representing fullerene molecular structure are called fullerene graphs where nodes and edges correspond to atoms and bonds of a molecule respectively.
Do{\v{s}}li{\'c} recommended the use of bipartivity measures in this context based on observing strong correlations between a bipartivity measure and several fullerene stability indicators. The correlations were evaluated on a set of eight experimentally verified fullerenes (produced in bulk quantities) with atom counts ranging between 60 and 84 \cite{doslic2005bipartivity}. 
Do{\v{s}}li{\'c} suggested using the \textit{spectral network bipartivity measure}, denoted by $\beta(G)$, which was originally proposed by Estrada et al.\ 
\cite{estrada2005}. 
This measure equals the proportion of even-length to total closed walks as formulated in 
\eqref{4eq5}
in which $\lambda_j$ ranges over eigenvalues of $|\textbf{A}|$ (the entrywise absolute value of adjacency matrix $\textbf{A}$). Note that $\beta(G)$ ranges between $0.5$ and $1$ and greater values represent more bipartivity. 
\begin{equation}\label{4eq5}
\beta(G)=\frac{\sum_{j=1}^{n} \cosh {\lambda_j}}{\sum_{j=1}^{n} e^{\lambda_j}}
\end{equation}
Two years later, Do{\v{s}}li{\'c} and Vuki{\v{c}}evi{\'c} suggested using the \textit{bipartite edge frustration} as a more intuitive measure of bipartivity to investigate the stability of fullerenes
\cite{doslic_computing_2007}. This measure equals the minimum number of edges that must be removed to make the network bipartite \cite{holme2003,yannakakis1981edge} and is closely related to the frustration index of signed graphs. Subfigure \ref{fig10a} shows a graph that is made bipartite in Subfigure \ref{fig10b} after removing 24 edges.
\FloatBarrier
Do{\v{s}}li{\'c} and Vuki{\v{c}}evi{\'c} have observed no strong correlation between the bipartite edge frustration and $\beta(G)$ 
\cite{doslic_computing_2007}.
However, both measures have performed well in detecting the most stable fullerenes among all isomers with 60 and 70 atoms 
\cite{doslic_computing_2007}. 
More recently, Estrada et al.\  \cite{estrada2016} 
proposed \textit{spectral bipartivity index}, denoted as $b_s(G)$ and formulated in \eqref{4eq6},
as a bipartivity measure with computational advantages over $\beta(G)$. Note that $b_s(G)$ ranges between $0$ and $1$ and greater values represent more bipartivity.
\begin{equation}\label{4eq6}
b_s(G)=\frac{\sum_{j=1}^{n} e^{-\lambda_j}}{\sum_{j=1}^{n} e^{\lambda_j}}=\frac{\Tr(e^{-\textbf{A}})}{\Tr(e^{\textbf{A}})}
\end{equation}
There are two other frustration-based measures of bipartivity suggested in
\cite{holme2003}.
Both measures are based on approximating $L(G)$ and we do not consider them.

\subsubsection{Relevance}
The bipartite edge frustration of a graph is equal to the frustration index of the signed graph, $G$, obtained by declaring all edges of the fullerene graph to be negative. Using this analogy, we provide some results on the bipartivity of large fullerene graphs. 
According to Do{\v{s}}li{\'c} et al., a motivation for using the bipartite edge frustration is exploring the range of atom counts for which there are no confirmed stable isomers yet 
\cite{doslic_computing_2007}.
The least bipartite fullerene graphs represent the most stable fullerene isomers \cite{doslic2005bipartivity, doslic_computing_2007}.
Therefore, lower bipartivity (smaller values of $\beta(G)$, $b_s(G)$, and $F(G)$) can be interpreted as higher stability.

\subsubsection{Datasets}
We use the binary linear model in Eq.\ \eqref{eq4} to compute the bipartite edge frustration of several fullerene graphs with atom count ranging from 180 to 6000. The fullerene graph of C240 (molecule with 240 carbon atoms) and its bipartite subgraph are visualised in Subfigures \ref{fig10a} -- \ref{fig10b} followed by C2160 and C6000 fullerene graphs in Subfigures \ref{fig10c} -- \ref{fig10d}.

Among the 14 fullerene graphs we consider are the \textit{icosahedral fullerenes} that have the structure of a truncated icosahedron. It is conjectured that this family of fullerenes has the highest chemical stability among all fullerenes with $n$ atoms \cite{doslic_computing_2007, faria2012odd}.

\subsubsection{Results}
The values of bipartivity measures for 14 fullerene graphs are computed in Table~\ref{tab4} where we have also provided the normalised frustration index, $F(G)=1-2L(G)/m$, to compare the bipartivity of fullerenes with different atom counts. The closeness of $F(G)$, $\beta(G)$, and $b_s(G)$ values to 1 are consistent with fullerene graphs being almost bipartite (recall that these three measures take value 1 for a bipartite graph) \cite{doslic2005bipartivity}.
\begin{table}[ht]
	\centering
	\caption{Bipartivity measures computed for a range of large fullerene graphs}
	\label{tab4}
	\begin{tabular}{llllll}
	\hline
	Fullerene graph & $m$  & $L(G)$ & $F(G)$  & $\beta(G)$ & $b_s(G)$ \\ \hline
	C180            & 270  & 18     & 0.86667 & 0.99765    & 0.99529  \\
	C240$^\dagger$  & 360  & 24     & 0.86667 & 0.99823    & 0.99647  \\
	C260            & 390  & 24     & 0.87692 & 0.99837    & 0.99674  \\
	C320            & 480  & 24     & 0.9     & 0.99867    & 0.99735  \\
	C500            & 750  & 30     & 0.92    & 0.99915    & 0.99830  \\
	C540$^\dagger$  & 810  & 36     & 0.91111 & 0.99921    & 0.99843  \\
	C720            & 1080 & 36     & 0.93333 & 0.99941    & 0.99882  \\
	C960$^\dagger$  & 1440 & 48     & 0.93333 & 0.99956    & 0.99912  \\
	C1500$^\dagger$ & 2250 & 60     & 0.94667 & 0.99972    & 0.99943  \\
	C2160$^\dagger$ & 3240 & 72     & 0.95556 & 0.99980    & 0.99961  \\
	C2940$^\dagger$ & 4410 & 84     & 0.96190 & 0.99986    & 0.99971  \\
	C3840$^\dagger$ & 5760 & 96     & 0.96667 & 0.99989    & 0.99978  \\
	C4860$^\dagger$ & 7290 & 108    & 0.97037 & 0.99991    & 0.99983  \\
	C6000$^\dagger$ & 9000 & 120    & 0.97333 & 0.99993    & 0.99986  \\ \hline
		\multicolumn{4}{l}{$\dagger$ icosahedral fullerene}
	\end{tabular}
\end{table}

Both spectral measures provide a monotone increase in the bipartivity values with respect to increase in atom count. However, $F(G)$ seems to provide distinctive values for icosahedral fullerenes. In particular for this set of fullerenes, we observe $F(C240)\not>F(C180)$, $F(C540)\not>F(C500)$, and $F(C960)\not>F(C720)$ which are consistent with the conjecture that icosahedral fullerenes are the most stable isomers \cite{doslic_computing_2007, faria2012odd}. 
\subsubsection{Computations}

Computing the bipartite edge frustration of a graph in general is computationally intractable and heuristic and approximation methods are often used instead \cite{holme2003}. For bipartite edge frustration of fullerene graphs which are planar; however, a polynomial time algorithm of complexity $\mathcal{O}(n^3)$ exists \cite{doslic_computing_2007}. Previous works suggest that this algorithm cannot process graphs as large as $n=240$ \cite{doslic_computing_2007}. Our computations for obtaining $L(G)$ of fullerene graphs with 180--2940 atoms take from split second to a few minutes. The solve times for computing $L(G)$ of C3840, C4860, and C6000 are 29.8, 68.1, and 97.5 minutes respectively.

As indicated by Table~\ref{tab4} results, the optimisation-based method in \eqref{eq4} 
allows computing frustration-based measures of bipartivity in the range of atom counts for which there are no experimentally verified stable isomers yet. The performance of frustration-based fullerene stability indicators requires further research that is beyond our discussion in this paper.

\subsection{Ising models with $\pm1$ interactions}\label{s:d6}

Closely related to the frustration index of signed graphs, are the ground-state properties of Ising models. The most simple and standard form of Ising models represents patterns of atomic magnets based on interactions among spins and their nearest neighbours. A key objective in Ising models with $\pm1$ interactions is finding the spin configurations with the minimum energy \cite{friedli2017statistical}. The standard nearest-neighbour Ising model with $\pm1$ interactions and no external magnetic field is explained in what follows.

Each spin is connected to its neighbours in a grid-shaped structure. Two connected spins have either an aligned or an unaligned coupling. The positive (negative) interaction between two spins represents a coupling constant of $J_{ij}=+1$ ($J_{ij}=-1$) alternatively called matched (mismatched) coupling. Under another terminology from physics, positive and negative edges are referred to as ferromagnetic bonds and anti-ferromagnetic bonds respectively \cite{zaslavsky2017}. Each spin can either take an upward or a downward configuration. We discuss finding a spin configuration for a given set of fixed coupling constants that minimises an energy function \cite{friedli2017statistical}.

Frustration arises if and only if a matched (mismatched) coupling has different (same) spin configurations on the endpoints. The energy of a spin configuration is calculated based on the Hamiltonian function: $H=-\sum_{ij} J_{ij} s_i s_j$ in which the sum $\sum_{ij}$ is over all the coupled spins. Note that $J_{ij}$ represent the couplings limited to $\pm1$ in the Ising model with the type of interactions relevant to this study. $s_1, s_2, \dots , s_n$ are the decision variables that take values $+1$ or $-1$ and represent upward/downward spin configurations. The Hamiltonian function of these Ising models is very similar to the energy function in \cite{facchetti_computing_2011} that is also used in other studies \cite{iacono_determining_2010, esmailian_mesoscopic_2014, ma_memetic_2015, ma_decomposition-based_2017, Wang2016}. The problem of minimising $H$ over all possible spin configurations is NP-hard for many structures \cite{liers2004computing}.

\subsubsection{Relevance}

In order to make a connection between Ising models and signed graphs, we represent spins by nodes and spin configurations by node colours. Signs on the edges represent coupling constants where matched and mismatched couplings between spins are modelled as positive and negative edges respectively.

If $X^*$ represents the optimal colouring leading to $L(G)$ for a given signed graph, the minimum value of the corresponding Hamiltonian function can be calculated by $H(X^*)=- \sum_{i,j} a_{ij}(2x_{i}-1)(2x_{j}-1)$. The minimum value of $H$ is obtained based on the optimal spin configuration associated with $X^*$. Alternatively, one may consider the fact that frustrated edges and non-frustrated (satisfied) edges contribute values $1$ and $-1$ to the Hamiltonian function respectively. For an Ising model with $m$ edges, this gives $H(X^*)=
 2L(G)-m$ as the optimal Hamiltonian function value.
 
\subsubsection{Datasets}

\begin{figure}[ht]
	\centering
	\subfloat[2D Ising model $50\times50$]{\includegraphics[height=1.7in]{fig12a-eps-converted-to.pdf}%
		\label{fig11a}} 
	\hfil
	\centering
	\subfloat[3D Ising model $10\times10\times10$]{\includegraphics[height=1.9in]{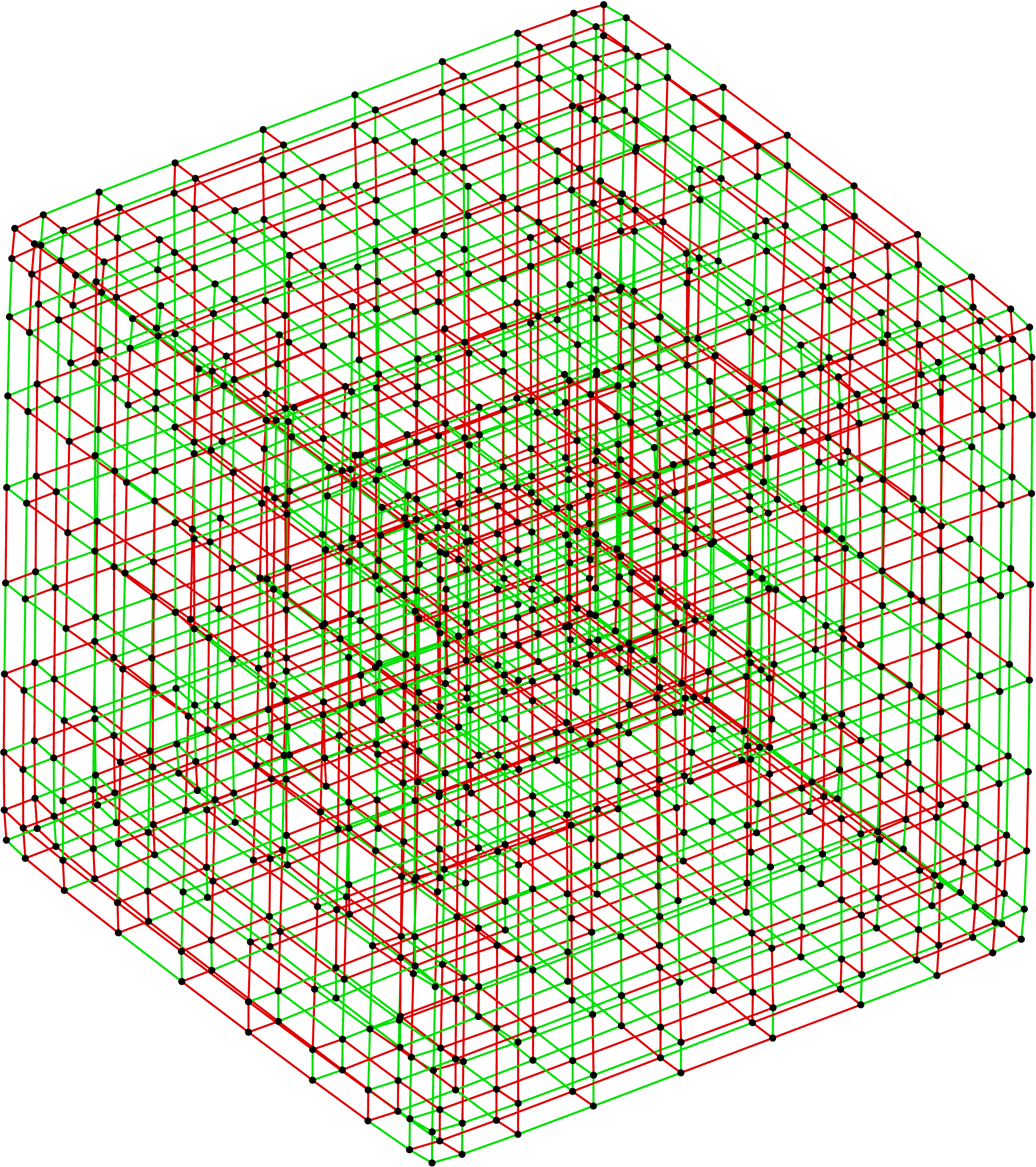}%
		\label{fig11b}} 
	\hfil
	\caption{Signed graphs with two and three dimensional structure representing simple Ising models with $50\%$ unaligned couplings 
	}
	\label{fig11}
\end{figure}

\begin{figure}[ht]
	\subfloat[Dimension 4]{\includegraphics[width=0.22\textwidth]{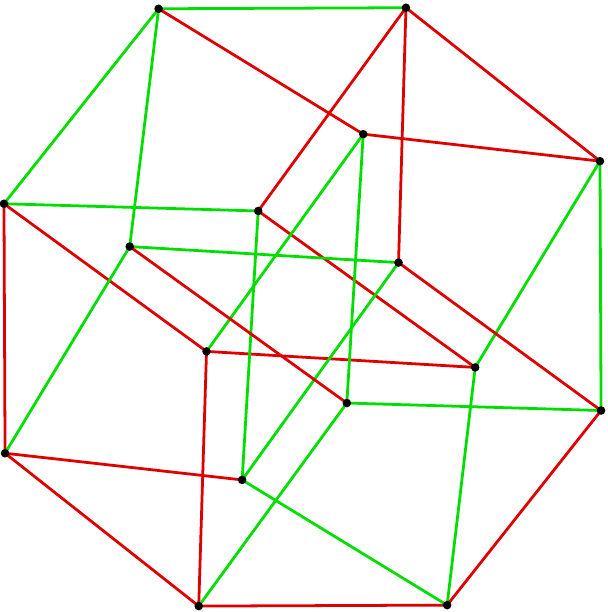}%
		\label{fig12a}} 
	\hfil
	\subfloat[Dimension 5]{\includegraphics[width=0.22\textwidth]{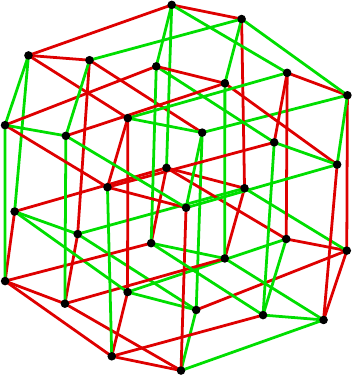}%
		\label{fig12b}} 
	\hfil
	\subfloat[Dimension 6]{\includegraphics[width=0.22\textwidth]{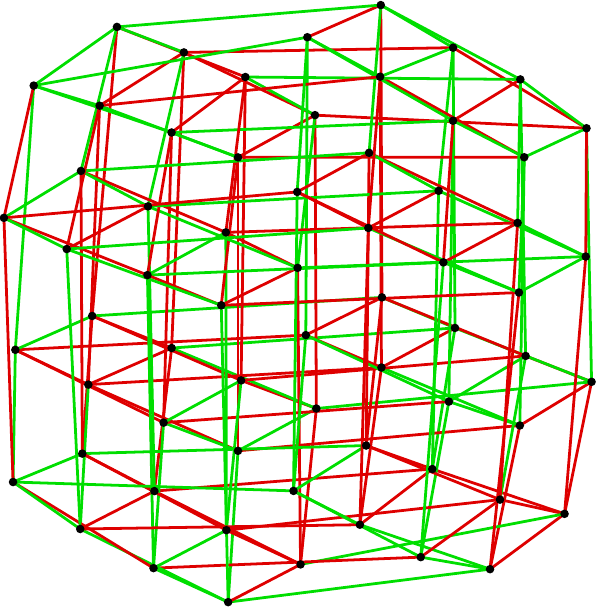}%
		\label{fig12c}} 
	\hfil
	\subfloat[Dimension 7]{\includegraphics[width=0.22\textwidth]{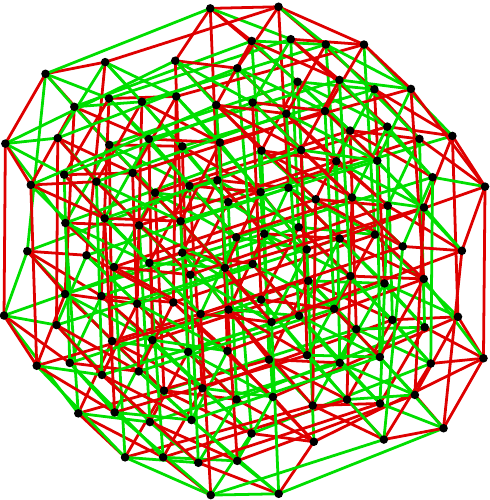}%
		\label{fig12d}} 
	\caption{Four signed graphs with hypercube structure representing more structurally complex Ising models with $50\%$ unaligned couplings 
}
	\label{fig12}
\end{figure}

We use the binary linear model in Eq.\ \eqref{eq4} to compute the frustration index in Ising models of various grid size and dimension for several 2D and 3D grid structures as well as hypercubes. Figure~\ref{fig11} illustrates a 2D and a 3D Ising model with $50\%$ unaligned couplings. Four hypercubes of dimension 4--7 with $50\%$ unaligned couplings are visualised in Figure~\ref{fig12}.

For each Ising model, we generate 10 grids and randomly assign $\pm1$ to the edges to achieve the pre-defined proportion of negative edges based on our experiment settings. For each Ising model with a specific dimension (Dim.) and grid size (Gri.), we consider three experiment settings with $m^-/m \in \{ 25\%, 50\%, 75\%\}$.

\subsubsection{Results}

Table \ref{tab5} provides results on Ising models with a fixed dimension and grid size in each row. The mean and standard deviation of the frustration index values and the average solve time (in seconds) for each experiment setting are provided in Table~\ref{tab5}.

\begin{table}
	\centering
	\caption{Frustration index values (and average solve times in seconds) for several Ising models}
	\label{tab5}
	\begingroup
	\renewcommand{\arraystretch}{1.5}
	\begin{tabular}{p{0.5cm}p{0.8cm}p{2.5cm}p{2.5cm}p{2.5cm}}
		\hline
		&										       & $m^-/m=25\%$        & $m^-/m=50\%$       & $m^-/m=75\%$       \\ \cline{3-5} 
		{Dim., \quad Gri.} 							    &      {$n$, \quad \quad$m$}                   & $L(G)$ mean$\pm$SD (solve time)   & $L(G)$ mean$\pm$SD (solve time)  & $L(G)$ mean$\pm$SD (solve time)\\ \hline 
		2,    \quad                  50                      & 2500, 4900                                   & 691.1$\pm$12 \quad (28.2)  & 720.9$\pm$9.2 \quad (31.9) & 687.7$\pm$10.5 \quad (42.3)  \\ 
		2,                      100                     & 10000, 19800                                 & 2814.1$\pm$16 \quad (2452.3)     & 2938.2$\pm$22.3  \quad (2660.8)  & 2802.5$\pm$24.7 \quad (4685.2) \\ 
		2,                      150                     & 22500, 44700                                 & 6416.1$\pm$25.9  \quad (5256.2)  & 6698.5$\pm$55.3  \quad (5002.0)  & 6396.3$\pm$41.7 \quad (6761.0) \\ 
		2,                      200                     & 40000, 79600                                 & 11449$\pm$47     \quad (13140.6)  & 11930.3$\pm$58.9 \quad (12943.7)  & 11411.8$\pm$46  \quad (22720.4) \\ \hline 
		3,  \quad                    5                       & 125,         300                             & 51.4$\pm$1.7     \quad \quad (0.1)  & 52.4$\pm$2.5     \quad \quad (0.1)  & 51$\pm$3.2     \quad \quad (0.1)  \\  
		3,  \quad                    10                      & 1000,        2700                            & 491.5$\pm$7.5    \quad (82.9)  & 509.1$\pm$4      \quad (539.0)  & 491.6$\pm$7    \quad (96.4)  \\  
		3,   \quad                   15                      & 3375,       9450                             & 1762.1$\pm$16.1  \quad (8488.2)  & 1839.1$\pm$10.4  \quad (21384.3)  & 1761.1$\pm$14.2 \quad (9244.1) \\ \hline
		4,   \quad                   2                       & 16,             32                           & 5.6$\pm$0.8   \quad   \quad (0.1)  & 4.8$\pm$1   \quad   \quad    \quad (0.1)  & 5.6$\pm$0.8  \quad   \quad (0.1)  \\  
		5,   \quad                   2                       & 32,            80                            & 14.5$\pm$1.1   \quad   \quad (0.1)  & 15$\pm$1.2   \quad     \quad (0.1)  & 15$\pm$1.2   \quad   \quad (0.1)  \\  
		6,   \quad                   2                       & 64,           192                            & 38.8$\pm$2    \quad    \quad (0.1)  & 41$\pm$1.6    \quad    \quad (0.1)  & 38$\pm$2.2   \quad   \quad (0.1)  \\  
		7,   \quad                   2                       & 128,          448                            & 94.6$\pm$3.1   \quad   \quad (0.6)  & 99.6$\pm$3.2   \quad   \quad (1.6)  & 96$\pm$2.4   \quad   \quad (1.1)  \\  
		8,   \quad                   2                       & 256,         1024                            & 232.4$\pm$3.7    \quad (206.1)  & 245.8$\pm$3.6    \quad (4742.2)  & 231$\pm$4.7    \quad (543.87)  \\ \hline
	\end{tabular}
	\endgroup
\end{table}
\FloatBarrier

The results in Table~\ref{tab5} show that in most cases the Ising model with $m^-/m = 50\%$ has the highest frustration index value (and therefore the highest optimal Hamiltonian value) among models with a fixed grid size and dimension. This can be explained by considering that in the structures investigated in Table~\ref{tab5} all cycles have an even length. Therefore a higher number of negative cycles (each containing at least one frustrated edge) is obtained when the number of positive and negative edges are equal.

\subsubsection{Computations}

Hartmann and collaborators have suggested efficient algorithms for computing the ground-state properties in 3-dimensional Ising models with 1000 nodes \cite{hartmann2015matrix} improving their previous contributions in 1-, 2-, and 3-dimensional \cite{hartmann2014exact,hartmann2011ground,hartmann2013information} Ising models. Recently, they have used a method for solving binary optimisation models to compute the ground state of 3-dimensional Ising models containing up to $268^3$ nodes \cite{hartmann2016revisiting}. While there are computational models for specialised Ising models based on the type of underlying structure \cite{hartmann2014exact,hartmann2011ground,hartmann2013information,hartmann2015matrix,hartmann2016revisiting}, the binary linear programming model in Eq.\ \eqref{eq4} can be used as a general purpose computational method for finding the ground state of Ising models with $\pm1$ interactions regardless of the underlying structure.


\section{Conclusion} \label{s:conclu}


In this study, the frustration index is used for analysing a wide range of signed networks from sociology and political science (Section \ref{s:d1}), biology (Section \ref{s:d2}), international relations (Section \ref{s:d3}), finance (Section \ref{s:d4}), and chemistry and physics (Section \ref{s:d56}) unifying the applications of a fundamental graph-theoretic measure. Our results contribute additional evidence that suggests many signed networks in sociology, biology, international relations, and finance exhibit a relatively low level of frustration which indicates that they are relatively close to the state of structural balance.

The numerical results also show the capabilities of the optimisation-based model in Eq.\ \ref{eq4} in making new computations possible for large-scale signed networks with up to $10^5$ edges. The mismatch between exact optimisation results we provided on social and biological networks in Sections \ref{s:d1} -- \ref{s:d2} and those in the literature \cite{dasgupta_algorithmic_2007, huffner_separator-based_2010, iacono_determining_2010, facchetti_computing_2011, ma_memetic_2015, ma_decomposition-based_2017} suggests the necessity of using accurate computational methods in analysing signed networks. This essential consideration is more evident from our results on international relations networks in Section~\ref{s:d3} where inaccurate computational methods in the literature \cite{patrick_doreian_structural_2015} have led to making a totally different inference with respect to the balance of signed international relations networks.

The current study provides extensive results on financial portfolio networks in Section~\ref{s:d4} confirming the observations of Harary et al.\ \cite{harary_signed_2002} on small portfolio networks being mostly in a totally balanced state. In Section~\ref{s:d56}, we extended the applications of the frustration index to a fullerene stability indicator in Subsection \ref{s:d5} and the Hamiltonian of Ising models in Subsection \ref{s:d6}. It is hoped that these discussions pave they way for using exact optimisation models for more efficient and reliable computational analysis of signed networks, fullerene graphs and Ising models.


While this study provided an overview of the state-of-the-art of the numerical computations on signed graphs and the vast range of applications to which it can be applied, it is by no means an exhaustive survey on the applications of the frustration index. From a computational perspective, this and some other recent studies \cite{Giscard2016, giscard2016general} call for more advanced computational models that put larger networks within the reach of exact analysis.
As another future research direction, one may consider formulating edge-based measures of stability for directed signed networks based on theories involving directionality and signed ties \cite{leskovec_signed_2010,yap_why_2015}.

\section*{Acknowledgement}
The authors would like to show their gratitude to the Centre for eResearch at University of Auckland for providing the high performance computer and to Marco De La Pierre, Tomislav Do{\v{s}}li{\'c}, and the anonymous referees for valuable comments and discussions that have improved this paper.




\bibliographystyle{plain}
\bibliography{refs}

\end{document}